Rolling contact fatigue deformation mechanisms of nickel-rich nickel-titanium-hafnium alloys

Sean H. Mills[1], Christopher Dellacorte[2], Ronald D. Noebe[2]

Behnam Amin-Ahmadi,[1*] and Aaron P. Stebner[1,3*]

[1] Mechanical Engineering Department, Colorado School of Mines, Golden, CO 80401, USA

[2] NASA Glenn Research Center, Materials and Structures Division, Cleveland, OH 44135, USA

[3] Mechanical Engineering and Materials Science & Engineering, Georgia Institute of Technology, Atlanta, GA 30332 USA

**ABSTRACT**

The tribological performance and underlying deformation behavior of $Ni_{55}Ti_{45}$, $Ni_{54}Ti_{45}Hf_1$ and $Ni_{56}Ti_{36}Hf_8$ alloys were studied using rolling contact fatigue (RCF) testing and transmission electron microscopy (TEM). TEM results of the as-machined RCF rods, prepared using focus ion beam, revealed some damage very close to the surface. TEM results after initial RCF cycling showed that additional damage was mainly confined to deformation bands that propagated several microns into the sample. These bands formed via localized dislocation slip, possibly on multiple slip systems, within the B2 matrix and/or within transformed B19´ martensite phase under repeated applied contact stress. Further cycling of $Ni_{55}Ti_{45}$ and $Ni_{54}Ti_{45}Hf_1$ led to shearing and dissolution of the strengthening precipitates within the deformation bands, followed by formation of nanocrystalline grains and finally amorphization of the remaining matrix material within the bands. The $Ni_{56}Ti_{36}Hf_8$ alloy exhibited a notable increase in RCF performance and a smaller damage zone (1.5 µm) compared to the $Ni_{55}Ti_{45}$ and $Ni_{54}Ti_{45}Hf_1$ alloys (over 6 µm). This was attributed to the low fraction of B2 matrix phase (≤ 13 %) in the $Ni_{56}Ti_{36}Hf_8$ alloy, leading to formation of narrow deformation bands (< 11 nm) that were incapable of dissolving the much larger precipitates. Instead, the deformation bands were restricted to narrow channels between the dense cubic NiTiHf and H-phase precipitates. In contrast, broad deformation bands accompanied by shearing and eventual dissolution of the $Ni_4Ti_3$ precipitates were observed in the $Ni_{55}Ti_{45}$ and $Ni_{54}Ti_{45}Hf_1$ alloys due to the high area fractions of B2 matrix phase (~49 %).

* Corresponding Authors: baminahmadi@mines.edu, aaron.stebner@gatech.edu

1. **Introduction**

High hardness binary nickel titanium (NiTi) alloys, with Ni content in the range of 52 – 56 at.%, exhibit attributes for tooling, bearings, and other tribological applications that are equal to and even surpass the capabilities of tool steels and ceramic bearing materials [1–6]. Unique features of these alloys include corrosion resistance, nonmagnetic behavior and high hardness, while having a lower density compared to tool steels [7,8]. Ni-rich NiTi alloys can be tailored to have a combined high hardness and low effective modulus, which allows these alloys to accommodate large amounts of strain with little permanent deformation. This enables the material to resist denting damage [9, 10] while achieving comparable rolling/sliding performance to tool steels [3-5]. Furthermore, NiTi alloys can be lubricated, unlike Ti alloys, and have an even lower friction coefficient than tool steels under some oil lubrication conditions [4,11,12]. Therefore, very Ni-rich NiTi alloys are drawing attention for high-performance tribological applications such as rotating centrifuge bearings in the water recycling system and the environmental control system in the International Space Station [2].

$Ni_{55}Ti_{45}$ and $Ni_{54}Ti_{45}Hf_1$ alloys have been developed for tribological applications [1-3]. Recently, Khanlari et al. [13] reported that the $Ni_{54}Ti_{45}Hf_1$ alloy exhibited a slightly higher wear resistance compared to $Ni_{55}Ti_{45}$. During dry-sliding testing, more subsurface cracks were observed in the $Ni_{55}Ti_{45}$ alloy compared to $Ni_{54}Ti_{45}Hf_1$ [13]. This enhanced fatigue and wear behavior was attributed to the effect of Hf on improving the microstructural homogeneity and, specifically, the prevention of $Ni_3Ti$ precipitation [13]. To more accurately simulate the complex wear effects that these materials exhibit while in service, rolling contact fatigue (RCF) testing was conducted [1]. Under RCF testing, the $Ni_{55}Ti_{45}$ alloy experienced erratic fatigue life and spalling failures began to occur at high Hertzian contact stress levels (~ 1.7 GPa) [1]. The $Ni_{54}Ti_{45}Hf_1$ alloy matched or exceeded the performance of the $Ni_{55}Ti_{45}$ alloy, which was assumed to be due to fewer inclusions [1]. In work concurrent to this article, we showed that $Ni_{56}Ti_{36}Hf_8$ alloy in aged condition performed the best of the newly developed alloys and better than the best performances attained of the $Ni_{55}Ti_{45}$ and $Ni_{54}Ti_{45}Hf_1$ alloys []. Still, the deformation mechanisms that accommodate RCF loading, and ultimately give way to spall failures have not been examined in depth for any of these nickel-rich alloys.

High resolution TEM (HRTEM) analysis of an unaged, less nickel-rich $Ni_{51.5}Ti_{48.5}$ alloy subjected to sliding wear testing at ~ 1 GPa Hertzian contact stress levels revealed an amorphous layer at the wear surface, then a region that contained B2 austenite and B19' martensite nanocrystals together with randomly distributed amorphous bands was observed just beneath the fully amorphized material [14]. The nanocrystalline microstructure was attributed to grain fragmentation caused by severe strains during repeated sliding wear, while further testing continued to break the fragmented B2 and B19' crystals into finer particles while the density of dislocations continued to increase, eventually leading to complete amorphization [14]. Note that this mechanism was proposed for Ni-rich NiTi alloys that did not contain dense strengthening precipitates via aging treatments. However, the $Ni_{55}Ti_{45}$, $Ni_{54}Ti_{45}Hf_1$, and $Ni_{56}Ti_{36}Hf_8$ alloys more recently developed for tribological applications are all strengthened by dense networks of nanoprecipitates in their peak RCF performance conditions []. The presence of The effects of these precipitates on sub-surface deformation mechanisms in response to tribological testing of these alloys remains unknown.

Still, amorphizatized and nanocrystalline microstructures have also been exhibited binary NiTi alloys subjected to severe plastic deformation (SPD) [22—27]. Amorphization is localized to narrow bands, originating from increased dislocation density along preferred slip planes resulting from heavy deformation [16,19]. Furthermore, amorphization can be assisted by the formation stress-induced martensite (SIM) where transformation induced dislocations act as sites for dislocation multiplication [14,21,22]. Local amorphization also has been observed in a $Ni_{50.3}Ti_{41.2}Hf_{8.5}$ alloy [23] around edges of cracks where a high concentration of Ni was also reported. In this case, Ni diffusion and amorphization were promoted by local stress concentrations ahead of the crack tip. Hence, amorphization and the formation of nanocrystals is anticipated to play a role in the RCF

Additionally, the quasi-static tension and compression deformation mechanisms of $Ni_{55}Ti_{45}$ and $Ni_{54}Ti_{45}Hf_1$ alloys have been studied. The microstructure of the baseline $Ni_{55}Ti_{45}$ alloy after repeated testing at high loads has been studied by Benafan et al. [24] in samples that were loaded to ~1.5 GPa in tension and ~ 2.5 GPa in compression for 50 cycles. Although the material was able to withstand these high cyclic stresses, coarsening and coalescence of the $Ni_4Ti_3$ phase and the precipitation of unwanted $Ni_3Ti$ phase along grain boundaries was observed even though cycling was performed at room temperature. This phase transformation behavior can lead to fatigue degradation, wear failure, and localized corrosion. Casalena et al. [25] studied the microstructure of the $Ni_{54}Ti_{45}Hf_1$ alloy after 10 cycles loading to ~ 2 GPa in compression. They did not observe coarsening of the $Ni_4Ti_3$ precipitate phase, but did report the presence of fine laths of retained B19' martensite, which was determined to be the primary source of unrecovered strain during the first loading cycle. The existence of retained martensite at room temperature was attributed to the presence of interfacial dislocations that accommodate the elastic incompatibility between the single-variant martensite and B2 matrix [25].

In this study, we examine the microstructural deformation mechanisms of the $Ni_{55}Ti_{45}$, $Ni_{54}Ti_{45}Hf_1$, and $Ni_{56}Ti_{36}Hf_8$ alloys using advanced electron microscopy techniques, before and after both run-out and spall failure RCF tests, to understand the reasons for the improved RCF performance exhibited by the $Ni_{56}Ti_{36}Hf_8$ alloy. We compare the deformation behavior of the $Ni_4Ti_3$ precipitate strengthened $Ni_{55}Ti_{45}$ and $Ni_{54}Ti_{45}Hf_1$ alloys that have been developed for bearing applications over the last decade with the best performing of the NiTiHf alloys most recently developed for RCF performance, the $Ni_{56}Ti_{36}Hf_8$ alloy [26,27]. This new alloy exhibits enhanced hardness and RCF performance that is correlated with a unique precipitate-strengthened microstructure consisting of densely packed, interspersed nanoprecipitates of both H-phase and a newly identified "cubic NiTiHf" precipitate phase []. The process-structure-property relationships of the are compared and contrasted through consideration of the different pre and post deformation, failed and passed RCF microstructures - specifically precipitate chemistries and morphologies as well as confinement of damage. Furthermore, the sequence and interplay of RCF damage mechanisms in these precipitate-strengthened nickel-rich NiTi/NiTiHf alloys is proposed.

## 2. Experimental

### 2.1. Materials and processing

The $Ni_{54}Ti_{45}Hf_1$ and $Ni_{56}Ti_{36}Hf_8$ alloys (target compositions) were vacuum-induction-melted (VIM) using a graphite crucible and cast into ingots 30 mm in diameter and 600 mm long. The ingots were homogenized at 1050 °C for 24 h, then sealed inside mild steel cans and hot-extruded into 9.5 mm (3/8-in.) diameter rods at 1000 °C. Chemical analysis of the extruded rods using inductively coupled plasma atomic-emission spectroscopy (ICP-AES) confirmed the composition of the alloys. The extruded rods were turned on a lathe to slightly oversize dimensions for RCF testing or a target of 10.16 mm diameter x 83.82 mm length. The rods were heat treated to achieve the peak hardness condition of all previously known heat treatment studies on each alloy system []. Specifically, the $Ni_{56}Ti_{36}Hf_8$ alloy was solution annealed at 1050 ºC for 0.5 h followed by a rapid water quench ($SA_{WQ}$). From there it was pre-aged at 300 ºC for 12 h followed by aging at 550 ºC for 4 h. The $Ni_{54}Ti_{45}Hf_1$ alloy was $SA_{WQ}$ (1050ºC for 0.5 h) and then aged at 400ºC for 0.5 h.

The baseline $Ni_{55}Ti_{45}$ alloy was manufactured via a high temperature powder metallurgy (PM) process in which the pre-alloyed NiTi powder was hot isostatic pressed at a temperature above 1000 ºC into large cylindrical compacts, the compacts were further homogenized at ~942 ºC and machined into RCF rods [28–33]. The composition of the PM rods was confirmed using ICP-AES. The $Ni_{55}Ti_{45}$ alloy rods were vacuum encapsulated in quartz tubes under Ar and $SA_{WQ}$ (1050 °C for 0.5 h) followed by aging at 400 °C for 1 h to achieve the peak hardness condition of previous studies [Hornbuckle 2015 influence of Hf, Hornbuckle 2015 decomposition, Casalena

2018] [].

The hardened blanks of all three alloys were then centerless ground to near final dimensions, 9.576 to 9.601 mm diameter, with a surface finish of 3–10 µm average surface roughness ($R_a$). Finally, the ground rods were "super-polished," to a final diameter of 9.520 to 9.525 mm for testing. "Super-polishing" is a commercial abrasion technique in which asperity tips are mechanically removed to attain a roughness comparable to a bearing raceway, typically better than 0.05 µm $R_a$ [1]. The $R_a$ value is the arithmetic mean of vertical surface deviations and is the universal roughness parameter for general quality control [35].

*2.2.  Rolling contact fatigue testing*

A Delta Research Corporation three ball-on-rod RCF test rig was used for evaluation of the longevity of the three alloys under complex RCF conditions by simulating types of loading and fatigue conditions that bearing components experience while in service [ref Olver 2005 and Ringsberg 2001]. A more in-depth description of the setup and functionality of this particular RCF tester is provided by Glover [36]. The NiTi and NiTiHf rod specimens were tested in contact with three polished and hardened steel balls rotated at 3600 rpm. The balls used in the rolling contact fatigue study were grade-ten, standard 12.7 mm diameter bearing balls made from hardened (700 – 800 HV) AISI 52100 type tool steel and were replaced after every test [1]. The balls were evenly spaced around the circumference of each rod by a bronze retainer and eight to ten drops of Mil-J-7808-J turbine oil drip feed lubrication per minute was provided lubrication between contacting parts, while effectively removing debris, regulating temperature, and preventing corrosion throughout the duration of the test [35]. A piezoelectric accelerometer was set close to ambient vibration (low sensitivity setting ~ 250 µm/$s_2$) to monitor surface damage, such as pit or spall formation on the wear track. When damage to the wear track caused vibrations greater than the set vibration limit, the test was automatically stopped and the number of cycles to failure was recorded. Otherwise, tests were manually halted at "runout" condition which is defined as $1.7 \times 10^8$ cycles (~ 800 h of bearing operation). In RCF, the maximum normal loading stresses do not occur at the same time as the maximum shear loading stresses. In addition, while the shear stresses may reverse loading direction, the normal stresses remain in compression [37]. The multiaxial stress state between two contacting bodies (in this case, a sphere contacting a cylinder) is complex and can be best explained by Hertzian contact theory [37,38]. Using this model which factors in contacting part geometries and material stiffness, the springs loaded on the test heads were adjusted (more detailed description of RCF test calibration available in Section 2.4 of Mills PhD thesis [ref]) to apply target contact loads on the RCF rods [ref thesis]. Initial contact stress levels of 1.8 GPa were selected based on previous reports of RCF performance of $Ni_{55}Ti_{45}$ alloy [ref Dellacorte]. Upon each completed test, the contact stress was increased (after a runout event) by an increment 100 MPa to define the transition from predominantly runouts to failures for each alloy.

A step-wise load increase method was adopted for determining the contact stresses that each alloy could sustain, where the contact force known to attain run-out conditions from previous RCF testing of the $Ni_{55}Ti_{45}$ alloy (180 N) [] was initially used for 3 to 5 tests of all alloys, and then loads were increased in 10 N increments each time a runout condition was observed. Each time a failure was observed, the load for the next test was decreased by 10 N. 9 mm spacers were used to adjust the height of the test sample such that fresh wear tracks were created for each test. The run-out condition used for this accelerated testing was $1.7 \times 10^8$ cycles (~800 h).

*2.3. Compression and tension testing*

Monotonic tension and compression tests of the two NiTiHf alloys were performed to failure using a 125 kN capacity MTS Landmark servo-hydraulic load. Compression and tensile data for the baseline $Ni_{55}Ti_{45}$ alloy is from Benafan, et al, [17]. For compression testing of the NiTiHf samples, 2mm diameter x 4mm long cylindrical specimens were centerless ground and then cut to length via electrical discharge machining (EDM). The samples were compressed using displacement controlled movement of the crosshead at a rate of 0.04 mm/s until failure. 2-dimensional Digital image correlation (2D-DIC) analysis of images using Ncorr open source 2D-DIC software package [ref Ncorr] was implemented to collect accurate strain information. The platen surfaces were coated with lubricant to minimize friction-induced off-axis loading effects caused by interactions between the platen and sample surfaces.

For tensile testing of the NiTiHf alloys, round dogbone specimens with cylindrical 5.10 mm diameter by 15.25 mm long gage sections were turned from the hardened rods using an EDM-lathe (see Fig X of [] for sample drawing). MTS 647.10 servohydraulic grips fitted with ____ wedges and set to a grip pressure of ____ MPa were used to grip the samples. Tensile loading was applied via displacement control of the crosshead at a rate of ____ mm/s. How were strains measured/calculated? Same DIC procedure? Or extensometer? Note that of the two $Ni_{56}Ti_{36}Hf_8$ tensile sample blanks that were available from this lot of material, one broke during machining and the other broke when hydraulic pressure was applied to the grips, hence a successful tensile test was not completed.

*2.4. Electron microscopy*

Conventional bright field (BF), dark field (DF), selected area electron diffraction (SAED)

and high-resolution transmission electron microscopy (HRTEM) techniques were used within a FEI Talos TEM (FEG, 200 kV). Cross-sectional TEM foils were prepared from the RCF rods via focused ion beam (FIB) milling using a "lift-out" technique from the rod surface. Samples were extracted from untested regions of the RCF rods (box c of Fig. 1(a), with corresponding secondary electron image (SEI) via scanning electron microscopy (SEM) in Fig. 1(c)), wear tracks that were created during successful run-out tests (box d of Fig. 1(a) and Fig. 1(d)), and directly under spalled regions of failed wear tracks (box e of Fig. 1(a) and Fig. 1(e)) . A protective Pt layer was deposited on the surface using the electron beam (lower energy), and then another higher energy layer using the ion beam before FIB milling was performed. This method minimizes damage to the sample by the incident $Ga_+$ ions. These protective layers are shown as deposited on top of a sample surface in the BF-micrograph in Fig. 1 (b). Ion beams of 30 kV per 3 nA and 30 kV per 0.3 nA were then used for cutting the surface and early-stage milling, with a probe size of 81 and 33 nm, respectively. An ion beam of 5 kV per 14 pA was used for the final cleaning step.

Dislocation densities were measured by counting extra half planes within HRTEM images. For better visualization of the extra half planes, a mask was applied at first order g-vectors and the corresponding Inverse Fast Fourier Transform (IFFT) was generated to visualize one family of planes. This procedure was used for each pair of first order {110} spots present in the <111> and <100> zone FFT patterns which were taken from regions within HRTEM images identified as B2 parent phase. The dislocation density was calculated based on the number of extra half planes present in all IFFT images. The reported accuracy of the dislocation density measurements was calculated by determining the allowed spread of the selected region in the HRTEM image that did not change the number of dislocations.

As will be shown in Section 3, four main deformation features were observed within the microstructures: the formation of deformation bands, dissolution of the strengthening precipitates, amorphization, and B2 (sometimes together with B19′) nanocrystallization. To quantify the morphologies and distribution of the deformation bands, selected area electron diffraction (SAED) (200 nm aperture) and HRTEM were used to determine the depths beneath the sample surfaces to which the other features were observed. To determine the depths at which the material was mostly amorphized, the diffraction pattern in SAED mode was dragged from the sample edge to the interior of the sample orthogonal to the edge direction. During this step, the diffraction was continuously analyzed for changes in the pattern which allowed us to locate transitions in the structure at varied depths. For example, the diffraction pattern held at the edge of a deformed sample typically contained a prominent diffuse ring pattern (amorphous structure) and scanning farther into the sample, the diffuse ring pattern transitioned to noticeable spot ring patterns (nanocrystalline structure). The approximate location (distance from the edge of the sample) of the transition was recorded. To confirm this depth measurement, HRTEM micrographs were taken at the same locations at the edge of the sample where the transition was identified and FFT confirmed the nanocrystalline features that indicated the transition from amorphous to crystalline structure. Similar SAED pattern analysis and HRTEM image analysis was used to determine the depths to which precipitate dissolution commonly occurred and to determine the depths to which nanocrystals prevalently formed.

### 2.5. *Differential scanning calorimetry*

Differential scanning calorimetry (DSC) was performed using a TA Instruments Q100 V9.9 with heating and cooling rates of 10°C/min over a temperature range between −180 °C and 150 °C. However, using this method, thermal martensitic transformation was not observed for any of the alloys in the heat-treated conditions.

## 3. Results

### 3.1. Monotonic tension and compression behaviors

The compressive stress – strain responses are plotted in Fig. 2(a). The $Ni_{56}Ti_{36}Hf_8$ alloy exhibited superior uniaxial compression strength (3.4 GPa yield stress) and moderate strain (4.2%) before failure by fracturing into several pieces. As previously reported [24], the $Ni_{55}Ti_{45}$ alloy yielded at ~2.7 GPa and exhibited a compressive strain of 3.2 % prior to failing by buckling. The $Ni_{54}Ti_{45}Hf_1$ alloy exhibited superelastic-like behavior at a stress level of ~ 1GPa that is consistent with the stress-induced transformation behavior reported by Casalena et al. [25], and achieved a maximum compressive stress of ~2.8 GPa and strain of 8.8% prior to failure by fracturing cleanly into 2 pieces about a plane 45º to the loading axis; i.e., a plane of maximum shear stress.

The tensile stress-strain responses (Fig. 2(b)) show that the $Ni_{55}Ti_{45}$ alloy tested by Benafan at al. [24] achieved an ultimate tensile strength of 1.6 GPa prior and 2.3 % elongation to failure. The $Ni_{54}Ti_{45}Hf_1$ alloy exhibited a tensile elastic limit of ~ 800 MPa. It is unclear if the inelastic deformation is due to phase transformation or plasticity or both, as the sample fractured at a maximum tensile stress of 1.1 GPa and 3.1 % elongation to failure, and the mechanical response to this point did not show any hardening or other signatures typical of separated phase transformation vs. plastic deformation responses. As previously noted, the tensile behavior of the $Ni_{56}Ti_{36}Hf_8$ alloy was not determined due to samples breaking before testing.

### 3.2. Rolling contact fatigue performances

The RCF results for all three alloys are plotted in Fig. 3. The corresponding contact stress limit exhibited by the $Ni_{56}Ti_{36}Hf_8$ alloy (2.2 – 2.3 GPa) is about 25% and 15% higher than the stress limits exhibited by the $Ni_{55}Ti_{45}$ and $Ni_{54}Ti_{45}Hf_1$ alloys of 1.8 and 2.0 GPa, respectively. As previously reported [], this improvement in sustained contact stress corresponds with both increased hardness (769 HV vs. 688 and 677) and stiffness (E = 130 GPa vs. 104 GPa, 72 GPa, see Fig. 2a). However, the microstructural origins of the increased compressive and RCF strengths, hardness, and stiffness are unknown, hence we proceed to examine the microstructures before and after RCF testing in detail.

### 3.3. *Microstructures of undeformed, worn, and spalled material*

Table 1 summarizes the dislocation density, deformation band morphologies (width and spacing), as well as the depths below the wear surface within which precipitates dissolved, amorphization occurred, and nanocrystallization occurred for each alloy in each condition. The analysis of each microstructure is presented for each alloy in the following subsections.

#### 3.3.1 *Microstructures of the $Ni_{55}Ti_{45}$ alloy*

##### 3.3.1.1 Sample fabrication damage

Fig. **4** depicts the microstructure of the super-polished $Ni_{55}Ti_{45}$ sample at and below the surface of the rod. Analysis of these data indicates that there was a layer of damaged material near the surface due to machining and/or polishing of the rods. Considering the mechanics of the sample preparation steps, it is most likely that the majority of this damage occurred during the centerless grinding operation that preceded the super-polishing. Deformation bands with average width of 43 ± 24 nm and average spacing of 221 ± 114 nm (Table 1) are observed through analysis of the material within 2 µm of the surface in Fig. 4(a). Observations of deformation bands are sparser at depths greater than 2 µm.

The magnified BF-TEM micrograph (Fig. 4(b)) shows that the material nearest to the surface is the most heavily damaged; the density of deformation bands increases close to the surface. SAED patterns taken from several regions with a diameter of approximately 200 nm (indicated by white circles in Fig. 4(b)) are presented in Fig. 4 (I–III). The SAED pattern taken nearest to the surface (Fig. 4(I)) is composed of diffuse ring and spotty patterns, which correspond to the existence of amorphous phase and B2 nanocrystals, respectively. The existence of $Ni_4Ti_3$ phase throughout the undeformed, aged microstructure is expected based upon previous studies of the $Ni_{55}Ti_{45}$ alloy [24,39]. However, typical $Ni_4Ti_3$ superlattice reflections were not observed within this region, indicating that they have been dissolved during sample fabrication processes. The HRTEM micrograph and corresponding diffuse circular FFT pattern taken from a deformation band (Fig. 4(c,d)) confirms that the deformation bands close to the surface are amorphous. The HRTEM image, Fig. 4(c), shows that the interfaces of the deformation bands lie along $(110)_{B2}$ planes. Analysis of the HRTEM image also indicated that the dislocation density of the B2 matrix within 200 nm of the surface is $6.2 \pm 0.3 \times 10^{16}/m^2$ (Table 1).

However, the material 200 – 400 nm below the surface shows a weak diffuse ring pattern together with bright spots from a B2 crystal oriented along a $[012]_{B2}$ zone, an addition to some weaker spots in the SAED pattern (Fig. 4(II)). These changes in the diffraction pattern compared to Fig. 4(I) implies that the material is mostly crystalline, containing fewer nanocrystals and less amorphous phase than the material within the first 200 nm of the surface. Weak superlattice reflections along $\frac{1}{7}\langle 321 \rangle$ are also observed, confirming the existence of some $Ni_4Ti_3$ precipitates 200 – 400 nm beneath the surface, though not in the expected morphologies, indicating that some precipitate dissolution has occurred.

The SAED pattern taken 400 – 600 nm beneath the surface (Fig. 4(III)) indicates a single B2 grain containing well formed $Ni_4Ti_3$ precipitates, evidenced by a strong spot pattern formed by reflections of a B2 grained oriented along the $[012]_{B2}$ zone together with smaller super lattice reflections of a single variant of $Ni_4Ti_3$. The mottling due to strain contrasts in the higher magnification BF-TEM (Fig. 4(b)) is also typical of the expected precipitate morphology that results from the hardening heat treatment. Hence, while some deformation bands are present at depths greater than 400 nm beneath the surface, larger undamaged regions of material exists at these depths; that is, regions where the microstructure is predominantly large grained B2 crystals containing a dense distribution of $Ni_4Ti_3$ precipitates.

*3.3.1.2 Wear damage*

The lowest magnification BF-TEM micrograph of worn, passed $Ni_{55}Ti_{45}$ material (Fig. 5(a)) appears similar to the unworn material (Fig. 4(a)). However, analysis of the deformation bands shows that they coarsened due to the wear to $63 \pm 37$ nm wide and average spacings of $153 \pm 97$ nm (Table 1). Also, the deformation bands appear more prevalent at depths 2 µm to 3 µm beneath the surface than they were in the untested material.

SAED patterns taken at different depths beneath the surface, as indicated with circles in the higher magnification BF-TEM micrograph, Fig. 5(b), are presented in Fig. 5(I,II). The SAED pattern within 300 nm of the surface shows a spot pattern of a single grain oriented along the [013]$_{B2}$ zone (Fig. 5(I)) together with a strong diffuse ring pattern originating from amorphized material. HR-TEM analysis (not shown) indicated that the material within 100 nm of the surface was mostly amorphized. Weak random superlattice reflections in this pattern can be attributed to a low volume fraction of retained Ni$_4$Ti$_3$ precipitates, though most of them have dissolved. Furthermore, the measured dislocation density of the B2 matrix increased to $11 \pm 0.2 \times 10^{16}$ /m$^2$ (Table 1), nearly double that observed near the surface in the super-polished condition.

SAED patterns taken between regions I and II in Fig. 5b were very similar. Then, the SAED pattern taken 800 nm from the surface (Fig. 5(II)) showed a sudden change. A sharp spot pattern corresponding to a B2 crystal structure oriented along the [013]$_{B2}$ zone axis appeared together with bright, well-defined superlattice reflections originating from two variants of Ni$_4$Ti$_3$ precipitates. These diffraction pattern features are expected of undamaged material. Still, the additional presence of a weak diffuse ring pattern suggests the existence some amorphous phase at this depth. HRTEM analysis confirmed that the amorphous structure was within the deformation bands.

*3.3.1.3 Spall damage*

The BF-TEM micrograph taken of material extracted underneath a spalled region of a failed wear track of the Ni$_{55}$Ti$_{45}$ alloy shown in Fig. 6(a) clearly shows much greater damage than Figs. 4(a) and 5(a). The SAED pattern taken near the surface (Fig. 6(I)) consists of a diffuse ring pattern (indicative of amorphous phase) and spot ring patterns from nanocrystalline grains of B2 and B19' structure. A part of the diffuse ring (indicated by dashed circle in Fig. 6(I)) was used to image the DF-TEM micrograph shown in Fig. 6(b). In this case, most of the microstructure under the spalled surface, to a depth of 3 µm, was amorphized; however, some dark contrast from nanocrystals was also observed within this region. Moreover, it is clear from the DF-TEM micrograph, Fig. 6(b), that the deformation bands far into the sample were amorphous. The RCF damaged penetrated more than 6 µm into the sample (limitation of FIB sample size). The average width of the amorphous deformation bands was 92 ± 34 nm and average spacing was 112 ± 75 nm (Table 1) – they have coarsened substantially compared with the worn, but passed material. The magnified BF-TEM micrograph (Fig. 6(c)) taken from the region indicated by the white box in Fig. 6(a) and the SAED pattern taken close to the surface (Fig. 6(I)) indicates that the structure was mostly amorphous 300 nm into the sample, together with some nanocrystals. The HRTEM micrograph taken close to the surface and corresponding FFT shown in Fig. 6(d) confirm that some of the nanocrystals are of B19' martensite structure. Further HRTEM analysis of the B2 matrix near the surface showed that the dislocation density increased to $13.1 \pm 0.2 \times 10^{16}$ /m$^2$ (Table 1), which seems to be the limit that the matrix can tolerate before amorphization occurs.

A high density of deformation bands is still observed deeper into the sample than the mostly amorphized region, as indicated by white arrows in Fig. 6(c). In fact, the SAED pattern taken 2.1 – 2.6 µm from the surface (Fig. 6(II)) contains a weak diffuse ring and intense spot ring patterns, indicating that even several microns below the surface, the material is mostly B2 and B19' nanocrystals in between amorphous deformation bands. Superlattice reflections from Ni$_4$Ti$_3$ precipitates were not detected in SAED patterns acquired up to 4.5 µm into sample. The SAED pattern (Fig. 6(III)) taken 4.1 – 4.5 µm from the surface, however, shows an intense spot pattern oriented along the $[\bar{1}11]_{B2}$ zone containing a weak diffuse ring pattern and superlattice reflections from Ni$_4$Ti$_3$ precipitates. Thus, the severe damage under the spall penetrated to 4 µm depths. Below 4 µm depths, the material is mostly polycrystalline B2 matrix phase with the expected $Ni_4Ti_3$ precipitates, except for dispersed deformation bands of amorphous material (Fig. 6(b)). Since the microstructure was heavily distorted throughout the region where HRTEM was taken, measurements were taken directly from B2 structure nanocrystalline grains.

*3.3.2. Microstructures of the $Ni_{54}Ti_{45}Hf_1$ alloy*

*3.3.2.1. Sample fabrication damage*

The super-polished $Ni_{54}Ti_{45}Hf_1$ alloy microstructure close to the surface, shown in Fig. 7, exhibits similar signs of damage as that observed in the super-polished $Ni_{55}Ti_{45}$ alloy (Fig. 4). Low and high magnification BF-TEM micrographs (Fig. 7 (a,b)) show deformation bands with average width of 21 ± 8 nm with an average spacing of 64 ± 12 nm within the damaged region (Table 1). They are most dense within the material 1 µm beneath the surface of the sample. The SAED pattern taken near the surface (Fig. 7 (I)) shows a spot pattern of a single B2 grain oriented along [011]$_{B2}$ zone together with a diffuse ring pattern from amorphized material. $Ni_4Ti_3$ precipitates were not observed up to 350 nm beneath the surface of the sample. The HRTEM image (Fig. 7 (c)) and corresponding FFT (inset) taken close to the surface shows that the material within the deformation bands is mostly amorphous, but also contains B2 nanocrystals. The interface of the deformation band is zig-zagged, following (110)$_{B2}$ and $(\bar{1}0\bar{1})_{B2}$ planes. The dislocation density of the B2 matrix close to the surface is $3.6 \pm 0.3 \times 10^{16}$ /m$^2$, or about half of that observed in the super-polished $Ni_{55}Ti_{45}$ alloy (Table 1).

The SAED pattern taken further beneath the surface (250 – 450 nm, Fig. 7 (II)) shows a B2 crystalline structure oriented along the [011]$_{B2}$ zone together with a diffuse ring pattern from amorphouse material and $Ni_4Ti_3$ superlattice reflections. The diffuse ring pattern in Fig. 7(II) is not as bright and sharp as that in Fig. 7(I), indicating that less amorphous material is present at this depth beneath the surface. The existence of $Ni_4Ti_3$ phase throughout the aged and undeformed B2 microstructure is expected from previous studies of the $Ni_{54}Ti_{45}Hf_1$ alloy microstructure [25]. The weak diffraction signal from this phase, together with a change from unmottled to mottled contrast in the BF-TEM micrograph of Fig. 7(b) indicate that these precipitates were mostly dissolved up to ~ 350 nm beneath the surface due to deformation induced by sample preparation (again, most likely the centerless grinding operation).

*3.3.2.2. Wear damage*

Material beneath the surface of the worn, passed $Ni_{54}Ti_{45}Hf_1$ wear track is shown in the BF-TEM micrograph of Fig. 8(a). The average width of the deformation bands increased by 4X compared to the as-polished condition, to 85 ± 20 nm, while the average band spacing increased by more than 6X to 422 ± 232 nm (Table 1). Recall that in the $Ni_{55}Ti_{45}$ alloy, deformation bands were only observed to widen by ~ 20 nm while the space between them concomitantly decreased by ~ 65 nm; i.e., the changes in the deformation bands were well described as the existing bands coarsening due to the wear. In contrast, the changes in deformation band structure of the $Ni_{54}Ti_{45}Hf_1$ alloy exhibit significant coalescence, in addition to coarsening. The worn deformation band structures were most dense up to 3.5 µm beneath the surface, and some of them were observed to terminate at a grain boundary (white dashed line in Fig. 8(a)), though note that the contrast in the other grain is different – a few of them did transmit across the grain boundary as well.

The magnified BF-TEM micrograph (Fig. 8(b)) taken from the white box in Fig. 8(a) and corresponding SAED pattern taken from the surface (Fig. 8(I)) indicate that the dense amorphous bands within 200 nm of the surface of the super-polished material have deformed into mostly amorphous material that contains a sparse distribution of B2 nanocrystals, again evidenced by diffuse ring and spot patterns. However, from Fig. 8(b), it appears that the material within 50 nm of the wear surface was almost completely amorphized – contrast from nanocrystals is more prevalent 50 – 200 nm beneath the surface. $Ni_4Ti_3$ precipitates were not observed up to 250 nm into the sample. Following the same descriptions given in the previous analyses, the SAED pattern taken 250 – 450 nm from the surface (Fig. 8(II)), together with the dense mottling in the BF-TEM image (Fig. 8(b)) indicate that the material in between the deformation bands at depths greater than 250 nm below the surface is predominantly polycrystalline B2 matrix containing the expected distribution of $Ni_4Ti_3$ precipitates. Similar to the $Ni_{55}Ti_{45}$ alloy, the dislocation density of the B2 matrix close to the surface doubled due to wear, compared with the super-polished condition, only here to a lower overall density of $6.7 \pm 0.2 \times 10^{16}/m^2$ considering the lower density in the super-polished material (Table 1).

*3.3.2.3. Spall damage*

Electron microscopy observations of spalled $Ni_{54}Ti_{45}Hf_1$ (Fig. 9) indicates a microstructure similar to spalled $Ni_{55}Ti_{45}$ (Fig. 6). Dense distributions of deformation bands extended more than 7 µm beneath the spall surface (limitation of FIB sample size in the BF-TEM shown in Fig. 9(a)) in a morphology measured to be $79 \pm 53$ nm wide with average spacing of $93 \pm 46$ nm (Table 1). Relative to the worn, passed condition, the deformation bands are of the same average width, but greater variation in the widths was observed. However, they are much denser within the spalled material, as the spacing between them decreased by 4X. The magnified BF-TEM image (Fig. 9(b)) confirms that the deformation band density was higher close to the surface and further into the sample (1.5 – 7 µm) the bands were more dispersed.

The SAED pattern taken just beneath the surface (Fig. 9(I)) was predominantly a single diffuse ring, indicating the existence of a nearly completely amorphized layer of material within 200 nm of the spall surface. The SAED pattern taken 800 nm – 1 µm from the surface (Fig. 9(II)) contained diffuse rings and spot patterns corresponding to the coexistence of amorphous material and B2/ B19' nanocrystals. The SAED pattern taken 1.6 – 1.8 µm from the surface showed a much weaker diffuse ring with an intense spot pattern from a B2 crystal oriented along the $[\bar{1}11]_{B2}$ zone (Fig. 9(III)). Note that $Ni_4Ti_3$ precipitates were not observed anywhere in the foil, indicating a denuded precipitate zone that extended at least 7 µm beneath the spall surface. The dislocation density within the B2 nanocrystals that could be analyzed with HRTEM nearest to the surface was $9.9 \pm 0.2 \times 10^{16} /m^2$ (Table 1). This maximum measured dislocation density is less than that observed of the spalled $Ni_{55}Ti_{45}$ alloy, though sample preparation differences could also be a factor.

*3.3.3. Microstructures of the $Ni_{56}Ti_{36}Hf_8$ alloy*

*3.3.3.1. Sample fabrication damage*

Based on concurrent studies [] the microstructure of undamaged, aged $Ni_{56}Ti_{36}Hf_8$ material is expected to be a 3-phase mixture of ~ 10 to 20 nm sized H-phase precipitates and precipitates of a newly identified NiTiHf cubic phase densely packed together within a B2 matrix phase (e.g., Table 1 and Figs. 8-9 of []). The specimen FIB'd from the super-polished, unworn surface of a $Ni_{56}Ti_{36}Hf_8$ RCF sample is shown in Fig. 10. The lower resolution BF-TEM micrographs (Fig. 10(a-b)) show that much less damage occurred within this sample during fabrication relative to the $Ni_{54}Ti_{45}Hf_1$ (Fig.7(a-b)) and $Ni_{55}Ti_{45}$ (Fig. 4(a-b)) alloys. The expected mottled contrast in the BF-TEM images from the 3-phase nanostructured mixture is evident throughput the sample. Deformation bands are only observed within 300 nm of the surface. Comparatively, they are incredibly fine, with average width of 9 ± 3 nm (indicated by white arrows), and also sparser, with average spacing of 178 ± 86 nm (Table 1).

The SAED pattern in Fig. 10(I) taken from region I in Fig. 10(a) shows a spot pattern of a single B2 grain oriented along the [001]$_{B2}$ zone together with superlattice reflections originating from H-phase [40,41] and the cubic NiTiHf precipitate phase, confirming the expected microstructure exists throughout the sample [add a ref to the current paper describing this phase]. Consistent with other reports of undeformed, aged Ni$_{56}$Ti$_{36}$Hf$_{8}$, the average size of the cubic NiTiHf precipitates was measured to be 17 ± 3 nm and H-phase precipitates 25 ± 7 nm (length), 15 ± 2 nm (width). The average width of the narrow B2 channels is 9 ± 2 nm and the measured dislocation density within these channels is $4.1 ± 0.3 \times 10^{16}/m^2$. The weak intensity of the diffuse ring pattern in Fig. 10 (I) is consistent with the signal expected from the existence of amorphous material within the fine, sparse deformation bands. The HRTEM image (Fig. 10(c)) and corresponding FFT (inset) taken just beneath the surface confirm the amorphous structure of a 2 – 4 nm wide deformation band surrounded by crystalline nano precipitates and B2 matrix. However, other deformation bands that extended further beneath the surface were observed to be crystalline, of B2 structure.

*3.3.3.2. Wear damage*

The microstructure of worn Ni$_{56}$Ti$_{36}$Hf$_{8}$ material under a passed wear track of an RCF sample is shown in Fig. 11. The BF-TEM micrographs of Fig. 11(a,b) show narrow deformation bands, such as those indicated with white arrows in Fig. 11(b). The average width of the deformation bands was 10 ± 3 nm, their average spacing was 102 ± 36 nm within the material 500 nm beneath the surface. The SAED pattern taken at the surface (Fig. 11(I)) shows a spot pattern of a single B2 grain oriented along the $[\bar{1}11]_{B2}$ zone, a diffuse ring pattern from amorphous material, and weak superlattice reflections from H-phase and cubic NiTiHf precipitates very near the surface. The low intensity of the precipitates indicates that some precipitate dissolution occurred within the first several hundred nanometers of material beneath the wear surface. The HRTEM micrograph in Fig. 11(c), taken close to the surface, shows that amorphization occurred within a narrow (2 – 4 nm wide) deformation band that has interfaces that lie along $(01\bar{1})_{B2}$ planes.

300 – 500 nm beneath the surface, the SAED pattern (Fig. 11(II)) contains a very weak diffuse ring originating from the presence of a few amorphous bands and an intense spot pattern of a B2 crystal oriented along the $[\bar{1}11]_{B2}$ zone together with H-phase and cubic NiTiHf superlattice reflections. Similar to the other alloys, the dislocation density within the narrow (11 ± 5 nm) B2 channels approximately doubled relative to the super-polished condition, to $7.7 \pm 0.2 \times 10^{16}$/m$^2$. H-phase precipitates coarsened to 43 ± 17 nm (length) and 28 ± 8 nm (width) due to the wear; however, B2 channel sizes and cubic NiTiHf precipitate sizes did not change. This result implies that the H-phase precipitates tend to coalesce as a result of deformation.

*3.3.3.3. Spall damage*

The spalled microstructure of the Ni$_{56}$Ti$_{36}$Hf$_8$ alloy is shown in Fig. 12. The BF-TEM micrographs of Fig. 12(a,b) indicate that most of the damage occurred within 1.5 µm beneath the surface, in contrast to the other alloys where significant damage was observed throughout the entirety of the spalled samples (Ni$_{55}$Ti$_{45}$ (Fig. 6(a)) and Ni$_{54}$Ti$_{45}$Hf$_1$ (Fig. 9(a))). Still, the spalled Ni$_{56}$Ti$_{36}$Hf$_8$ (Fig. 12(a,b)) is signficantly more damaged than the untested (Figs. 10(a,b)) and worn-but-not-failed (Figs. 11(a,b)) material. Narrow deformation bands within the damaged region have an average width of 10 ± 5 nm and are spaced 64 ± 18 nm apart. The diffuse ring and spots of the SAED pattern taken at the surface (Fig. 12(I)) indicate that just beneath the surface, the material structure is B2 and B19' nanocrystals grains within an amorphous phase. The amorphous surface layer was confirmed by the maze-like pattern in the HRTEM micrograph (Fig. 12(c)) and corresponding FFT pattern (inset). This mostly amorphous layer extended 100 nm beneath the surface.

200 – 400 nm beneath the surface, the diffuse ring and more intense B2/B19' spots of the SAED pattern (Fig. **12**(II)) indicate a higher volume fraction of nanocrystals exist together with the amorphous phase. Faint superlattice reflections (e.g., those marked with white arrowheads) corresponding to precipitate phases are also observed, indicating that precipitate dissolution occurred, but not completely. 700 – 900 nm beneath the spall surface fewer amorphous bands and more/larger H-phase and cubic NiTiHf precipitates were observed, as evidenced by the SAED pattern (Fig. **12**(III)) and the HRTEM micrograph and corresponding FFT (Fig. **12**(d)). Similar to the NiTiHf alloy, the dislocation density within the B2 material near to the surface, just beneath the amorphous layer, has increased to $9.2 \pm 0.2 \times 10_{16}$ /m$_2$.

## 4. Discussion

HRTEM analysis throughout this study has demonstrated that, regardless of composition and the type of damage, the interfaces of deformation bands follow $\{110\}_{B2}$ planes (Figs. 4(c), 7(c), 10(c), 11(c), 12(d)). This interface geometry excludes B2 twinning as a possible deformation mode to create the banded structures, since $\{112\}_{B2}$, $\{113\}_{B2}$, and $\{114\}_{B2}$ deformation twins are known to form in heavily deformed NiTi alloys [], but $\{110\}_{B2}$ twins have not been observed and are not geometrically plausible when twinning calculations [] are made for these alloys.

A likely mechanism for nucleating the deformation bands is localized dislocation slip within the B2 matrix. When the contact stress during RCF testing exceeds the yield strength of the samples, local plastic deformation in the form of dislocation slip occurs along preferred slip systems. $\{110\}_{B2}$ are common slip planes for dislocation slip in B2 alloys including NiTi [43], and such slip is consistent with the orientations of the deformation band interfaces within the B2 matrix across many alloys and deformation conditions. Moreover, the zig-zag nature of the interface of the deformation band shown in Fig. 7(c), implies that the activation of multiple $\{110\}_{B2}$ slip planes under RCF loading, or even cross-slip.

The other deformation mechanism that could also explain the formation of the deformation bands is martensitic transformation. Despite the absence of thermally induced martensitic transformation in DSC tests performed to -180 ºC, but consistent with in situ observations of stress induced martensitic transformation in the mechanical responses of the $Ni_{55}Ti_{45}$ and $Ni_{54}Ti_{45}Hf_1$ alloys [], B19' martensite nanocrystals were observed in several deformed states of all 3 alloys (Fig. 6, Fig. 9, Fig. 12), providing evidence that martensite was stress-induced during RCF testing. $\{110\}_{B2}$ are able to serve as geometrically compatible transformation interfaces for NiTi alloys [22,44]. Furthermore, recent observations of the $Ni_{54}Ti_{45}Hf_1$ alloy in the same aged condition after 10 sequential compressive loads showed bands of retained martensite confined by dislocations associated with $\{110\}_{B2}$ [], demonstrating that $\{110\}_{B2}$ dislocations (note, multiple slip directions are possible) can work cooperatively with martensitic transformation to stabilize deformation band structures in these nickel-rich, precipitate strengthened alloys.

Thus, to propose the means by which the deformation bands nucleate, coarsen, and coalesce, we consider the likely interplay of both dislocation slip and martensitic transformation. We also consider that in observing the untested, worn-but-not-failed, and spalled states of material at different depths below the rolling contact surfaces, gradients in combinations of damage mechanisms presented themselves ranging from mostly undamaged material that contained a few deformation bands of either amorphous or B2 crystalline structure microns beneath the surfaces of the untested materials all the way to fully amorphous material in the most heavily damaged regions, such as just below the spalled surfaces. At intermediate states of deformation, distinct deformation bands of martensite structure were not observed of any alloy, different from the Casalena et al. observations made after heavy uniaxial compressive deformation []. If even some the bands had nucleated and grown via martensitic transformation, and then been stabilized by slip, and then eventually more heavily deformed by slip and other mechanisms (which we discuss below) until they amorphized, we would expect to have observed deformation bands almost completely of martensitic structure in the same regions or slightly less deformed regions as we observed deformation bands of mostly B2 structure. However, we did not. In the RCF samples, the only observed martensite structures were small nanocrystals within mostly amorphous deformation bands and/or almost fully amorphized layers of material; i.e., the most heavily deformed regions, indicating that martensitic transformation occurred later in the sequence of deformation processes. Hence, we propose that in accommodating RCF damage, localization of B2 slip primarily served to nucleate and grow the deformation bands [].

In addition to the nucleation, growth, and coalescence of deformation bands, strengthening precipitates were observed to have dissolved, even in only moderately deformed regions of untested $Ni_{55}Ti_{45}$ and $Ni_{54}Ti_{45}Hf_1$ alloys (Figs. 4 and 7), and moreso in the more heavily deformed material states (Figs. 5-6, 8-9). Contrarily, precipitate dissolution was not evident of the untested $Ni_{56}Ti_{36}Hf_8$ alloy (Fig. 10), was limited in the worn material (Fig. 11), and was only observed to completion in the spalled material (Fig. 12). For the $Ni_{55}Ti_{45}$ and $Ni_{54}Ti_{45}Hf_1$ alloys, the undeformed structure of the material is large (tens of microns) B2 crystals that envelope distributions of $Ni_4Ti_3$ precipitates on the order of 20 – 40 nm in size and present in ~ 50% area fractions []. The undeformed $Ni_{56}Ti_{36}Hf_8$ structure in the aged condition, however, exhibits 10 – 25 nm sized H-phase precipitates in 54% area fraction and 10 – 20 nm sized cubic NiTiHf precipitates in 33% area fraction that confine B2 matrix ligaments that are 5 – 10 nm thick in between them. Only ~ 13% of the material is B2 crystalline by area fraction analysis of HR-TEM images []. When we now consider the morphologies of the deformation bands observed in the untested material (Table 1), in the $Ni_{55}Ti_{45}$ and $Ni_{54}Ti_{45}Hf_1$ alloys, they coarsen to ~ 10 – 60 nm thick and 10 – 30 nm thick, respectively – or to the same size as the strengthening precipitates. However, in the untested $Ni_{56}Ti_{36}Hf_8$ material, they are only 5 – 10 nm thick, smaller than the precipitates, but the same size as the B2 channels. Consistent with the aforementioned hypothesis that localization of slip in the B2 phase serves to nucleate the deformation bands, we observe that when the geometry of the B2 material available to the deformation bands is finer than the strengthening precipitates, the deformation bands seem to be confined by the precipitates phases and the precipitates do not dissolve until the material is heavily deformed. However, when there are larger regions of B2 material available to participate in slip localization, coarser deformation bands form, on the same order or larger than the strengthening precipitates, and the precipitates within the deformation bands are sheared by gliding dislocations along single or even multiple slip systems and eventually dissolve into the matrix.

The propensity for shearing and dissolution could also be affected by the chemistry of the precipitate phases. Here, H-phase precipitates were observed to initially coarsen within the moderately deformed regions of the $Ni_{56}Ti_{36}Hf_8$ alloy (Section 3.3.3.2), analogous to the previously observed coarsening of $Ni_3Ti$ precipitates within the $Ni_{55}Ti_{45}$ alloy after tensile deformation []. Certainly precipitate coarsening would inhibit shearing and dissolution. However, the cubic NiTiHf phase did not coarsen, but also did not shear and dissolve until heavier damage occurred. Hence, it is possible that the H-phase and cubic NiTiHf phases better resist deformation than $Ni_4Ti_3$ precipitates, though it is difficult to state with certainty due to the microstructural disparities of the morphologies of the undeformed materials.

The mechanism for precipitate dissolution, however, is rational. $Ni_4Ti_3$ precipitates within NiTi alloys are stiffer than the matrix [45], and H-phase and cubic NiTiHf phase are also expected to be since the bulk alloys harden when these precipitates are nucleated and grown in dense distributions. A small change of atomic spacing and arrangement of the precipitates by dislocation shearing can increase the free energy of precipitates several times more than that of the NiTi matrix. The process is repeated during each cycle, which reduces the size of particles to the point where they are unable to resist dislocation movement. Moreover, the preferred pathways for dislocation movement are better established during repetitive deformation and consequently, the pressure within the precipitates increases due to the release of surface tension, leading to an increase of the free energy of precipitates. When the accumulated free energy of the $Ni_4Ti_3$ precipitates is larger than that of the matrix, the sheared precipitates start to dissolve into the matrix [46]. This process results in the formation of precipitate-free material – first within the deformation bands where slip has heavily localized, and eventually throughout the material.

The mechanisms for the formation of B2 and B19′ nanocrystals is expected to be similar,

and to occur after precipitate dissolution since nanocrystals are only observed in material that is more heavily deformed than least deformed material that exhibits precipitate dissolution. In regard to forming B2 nanocrystals, continued RCF cycling increases the high dislocation density (Table 1). The mechanism for this process has been proposed by Zhang et al. [47] where dislocations of the same sign oriented on crossing slip planes form narrow channels, which become walls of individual dislocation cells within the larger grains or in this case the deformation band. Consequentially, the coarse grain is divided into a large number of sub-grains with a low-angle of misorientation [47]. Under further cycling, dislocation density is increased in the cell walls, which creates grain boundaries and the sub-grains are refined to nanocrystalline grains with a high-angle misorientation [19,48]. Eventually, additional plastic strain destroys the periodicity of the atomic arrangement in the nanocrystalline grains and causes amorphization [47]. Moreover, some of the amorphous deformation bands observed contain remnant nanocrystalline grains, which indicates this crystalline to amorphous transition is still occurring. The existence of some retained martensite nanocrystals in some of the same samples at room temperature within heavily deformed regions of material that did not fully amorphize suggests that the greatest deformation-induced stresses prior to amorphization, perhaps together with local chemistry changes such as local increases of Ni and/or Hf when the Ni- and/or Hf-rich precipitates dissolve, are sufficient to enable martensitic transformation of some of the nanocrystals via local elevations of the transformation temperatures through both chemical and Clausius-Clapeyron effects [,55,56].

Altogether, the proposed deformation mechanism sequence to accommodate RCF-induced damage in precipitate-strengthened, nickel-rich NiTi and NiTiHf alloys is schematically illustrated in Fig. 13. First, slip occurs within the B2 matrix. Then, localization of that slip nucleates deformation bands, and further concentration of the slip within the deformation bands coarsens them. Eventually, they consume the strengthening precipitates via shear and dissolution. Then, when the dislocation densities are high enough, cellular patterns of dislocations form nanocrystals, and localization of chemical and mechanical thermodynamic driving forces cause some of those nanocrystals to transform to martensite. Finally, full amorphization occurs, first within the bands, and then as the bands coalesce, eventually complete layers of amorphous material are created, leading to spall failures in the RCF tests when they can no longer accommodate anymore deformation.

Finally, note that it is known that inclusions, microstructural flaws, and voids can cause sporadic failures under RCF, which can limit performance [1]. Significant effort was spent investigating the failure surfaces at and about the spall sites using SEM and not a single failure could be traced to any kind of gross defect, such as an oxide, carbide or other inclusion or some kind of preexisting machining defect. Furthermore, while the statistics of the RCF study are somewhat limited, anomalous failures were not observed (Fig. 3). Still, these materials are known to contain Hf and Ti oxide particles [], and the binary alloys also brittle $Ni_3Ti$ particles [], all of which would certainly be likely candidates for failure initiation if seated directly on a contact surface. Thus, processing innovations to further reduce these undesired inclusions may help prevent erratic fatigue life.

## 5.    Conclusion

The aged Ni$_{56}$Ti$_{36}$Hf$_8$ alloy showed a tremendous resistance to RCF damage relative to the Ni$_{54}$Ti$_{45}$Hf$_1$ and Ni$_{55}$Ti$_{45}$ alloys, explaining the ability to tolerate greater contact stresses in RCF testing (Fig. 3). This fatigue resistance is primarily attributed to the geometries of the aged microstructures – in the improved alloy, the matrix material available for slip localization is of finer sizes than the strengthening nanoprecipitates; hence, the high phase fraction of strengthening precipitates of both H-phase and cubic NiTiHf phases promotes the confinement of damage that leads to eventual amorphization to occur only within very small (~ 10 nm wide) deformation bands until the material fails. Lower phase fractions of Ni$_4$Ti$_3$ precipitates allow for more B2 matrix material to participate in slip localization within the Ni$_{54}$Ti$_{45}$Hf$_1$ and Ni$_{55}$Ti$_{45}$ alloys; hence coarser deformation band structures form at sizes large enough to consume the strengthening Ni$_4$Ti$_3$ precipitates, and much more of the material is damaged more easily (e.g., simply by machining and polishing the RCF samples) to greater depths below the wear surfaces (tens of microns vs. several hundred nanometers). This conclusion is supported by the following summary of the main observations made in this work:

- Surface damage was detected in the as-machined, super-polished condition of the Ni$_{54}$Ti$_{45}$Hf$_1$ and Ni$_{55}$Ti$_{45}$ alloys, but minimal in the untested Ni$_{56}$Ti$_{36}$Hf$_8$ alloy. This result indicates that the machining steps during RCF rod fabrication, most likely the centerless grinding step, prior to super-polishing, can induce damage and significantly alter the microstructure very close to the surface. It is apparent that the Ni$_{55}$Ti$_{45}$ and Ni$_{54}$Ti$_{45}$Hf$_1$ alloys are more susceptible to this type of damage as the damaged regions extended 2 μm and 1μm into the specimens, respectively. Moreover, these two alloys contained precipitate free regions combined with a high density of deformation bands very close to the surface. This was not the case for the Ni$_{56}$Ti$_{36}$Hf$_8$ alloy, which exhibited only a very narrow damage

- zone (300 nm) and as-desired precipitate strengthened material throughout the as-prepared RCF samples.

- Inspection of the damage under the wear tracks of unfailed RCF runout ($1.7 \times 10^8$ cycles) specimens showed increased damage compared to the super-polished condition. The $Ni_{55}Ti_{45}$ and $Ni_{54}Ti_{45}Hf_1$ alloys both contained damaged regions that extended ~ 3.0 - 3.5 µm into the specimen. Both alloys experienced an expansion of the amorphous/nanocrystalline grain region and precipitate free zone depth into the sample. On the other hand, the $Ni_{56}Ti_{36}Hf_8$ alloy did not exhibit significant microstructural changes compared to the super-polished condition, i.e., precipitates close to the surface were not dissolved and the damaged region only extended to 500 nm.

- In sections near spalls in the $Ni_{55}Ti_{45}$ and $Ni_{54}Ti_{45}Hf_1$ alloys, the damaged regions consumed the entire samples. The regions closest to the surface, have become nearly fully amorphous and the dislocation density increased significantly. In the spalled $Ni_{56}Ti_{36}Hf_8$ sample, however, the damaged region extended only to 1.5 µm depth below the spalled surface. Most of the nanocrystalline grains that formed close to the surface had not yet amorphized, which indicates that this particular microstructure is more stable against damage even under higher contact stress levels.

- The mechanistic sequence for RCF damage in these alloys was determined, as summarized in Fig. 13, to be: First, slip occurs within the B2 matrix. Then, localization of that slip nucleates deformation bands, and further concentration of the slip within the deformation bands coarsens them. Eventually, they consume the strengthening precipitates via shear and dissolution. Then, when the dislocation densities are high enough, cellular patterns of dislocations form nanocrystals, and localization of chemical and mechanical thermodynamic driving forces cause some of those nanocrystals to transform to martensite. Finally, full amorphization occurs, first within the bands, and then as the bands coalesce, eventually complete layers of amorphous material are created, leading to spall failures in the RCF tests when they can no longer accommodate anymore deformation.

**Acknowledgements**

This work was conducted within the National Science Foundation (NSF) I/UCRC Center for Advanced Non-Ferrous Structural Alloys (CANFSA), which is a joint industry-university center between the Colorado School of Mines and Iowa State University. Additional support was provided for this work through the NASA Transformative Aeronautics Concepts Program (TACP), Transformational Tools & Technologies Project under the guidance of Othmane Benafan, Technical Lead for Shape Memory Alloys.

Tables

Table 1. Quantification of microstructure features.

| Surface Condition | Alloy | Dislocation density within B2 ($10^{16}$ m$^{-2}$) | Deformation band width (nm) | Deformation band spacing (nm) | Precipitate dissolution depth (nm) | Mostly amorphous depth (nm) | B2 and/or B19′ nanocrystals depth (nm) |
|---|---|---|---|---|---|---|---|
| Super-polished (untested) | Ni$_{55}$Ti$_{45}$ | 6.2 ± 0.3 | 43 ± 24 | 221 ± 114 | 400 | 0 | 200 |
| | Ni$_{54}$Ti$_{45}$Hf$_{1}$ | 3.6 ± 0.3 | 21 ± 8 | 64 ± 12 | 350 | 0 | 0 |
| | Ni$_{56}$Ti$_{36}$Hf$_{8}$ | 4.1 ± 0.3 | 9 ± 3 | 178 ± 86 | 0 | 0 | 0 |
| Worn (passed wear track) | Ni$_{55}$Ti$_{45}$ | 11.9 ± 0.2 | 63 ± 37 | 153 ± 97 | 800 | 100 | 800 |
| | Ni$_{54}$Ti$_{45}$Hf$_{1}$ | 6.7 ± 0.2 | 85 ± 20 | 422 ± 232 | 250 | 50 | 100 |
| | Ni$_{56}$Ti$_{36}$Hf$_{8}$ | 7.7 ± 0.2 | 10 ± 3 | 102 ± 36 | 0 | 0 | 0 |
| Spalled (failed wear track) | Ni$_{55}$Ti$_{45}$ | 13.1 ± 0.2 | 92 ± 34 | 112 ± 75 | 4500 | 300 | 6000 |
| | Ni$_{54}$Ti$_{45}$Hf$_{1}$ | 9.9 ± 0.2 | 79 ± 53 | 93 ± 46 | 2000 | 300 | 2500 |
| | Ni$_{56}$Ti$_{36}$Hf$_{8}$ | 9.2 ± 0.2 | 10 ± 5 | 64 ± 18 | 400 | 100 | 500 |

Figures

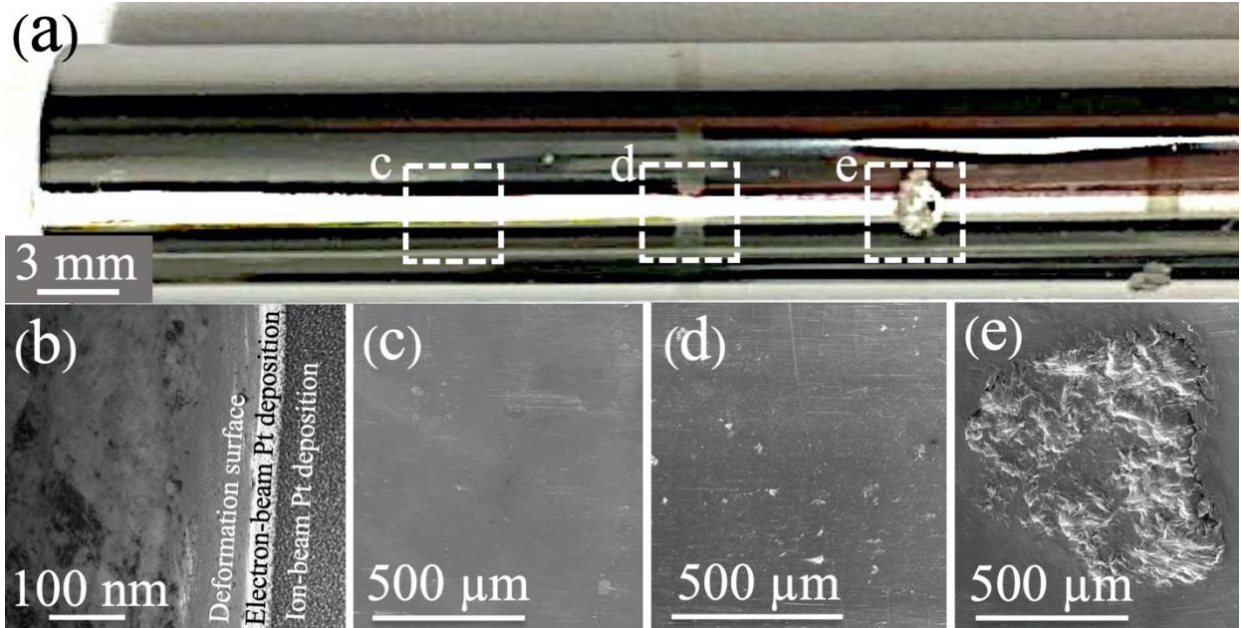

**Fig. 1.** (a) Typical rolling contact fatigue (RCF) test specimen after several tests, with as-polished regions of the surface such as the region labeled c, as well as multiple wear tracks such as the regions labeled d and e. (b) BF-TEM micrograph of a TEM foil prepared by FIB "lift-out" from the untested, super-polished rod surface showing the preserved deformation layer and the two Pt protective layers deposited using the electron-beam (lower energy) and Ga+ ion-beam (higher energy). SEM micrographs of (c) a super-polished region of the rod surface, (d) a wear track region of the surface after "run-out" was attained at $1.7 \times 10^8$ cycles, and (e) a spalled region of the surface within the wear track of a test that failed prior to $1.7 \times 10^8$ cycles.

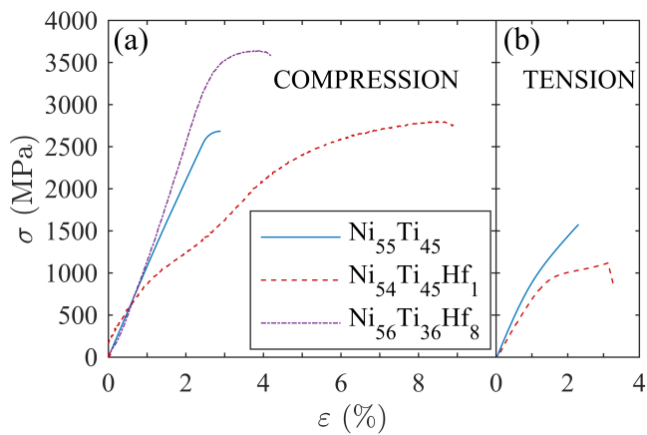

**Fig. 2.** (a) Compression and (b) Tension stress-strain behaviors of each alloy aged in the same condition as the RCF tests were performed. The Ni$_{56}$Ti$_{36}$Hf$_8$ tensile sample broke when gripped, and more material was not available to make more samples. The Ni$_{55}$Ti$_{45}$ data is replotted from [24].

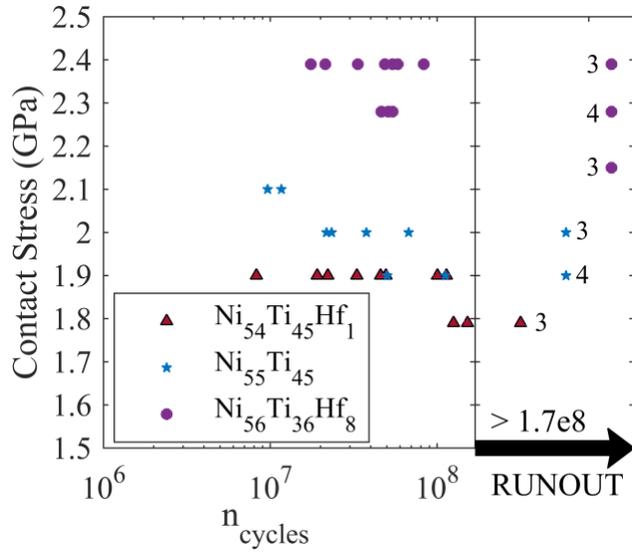

**Fig. 3.** Rolling contact fatigue results are plotted as a function of Hertzian contact stress (GPa) vs. cycles to failure (log10). The number of successful runout samples at each stress level are indicated next to the respective markers on the right-hand side.

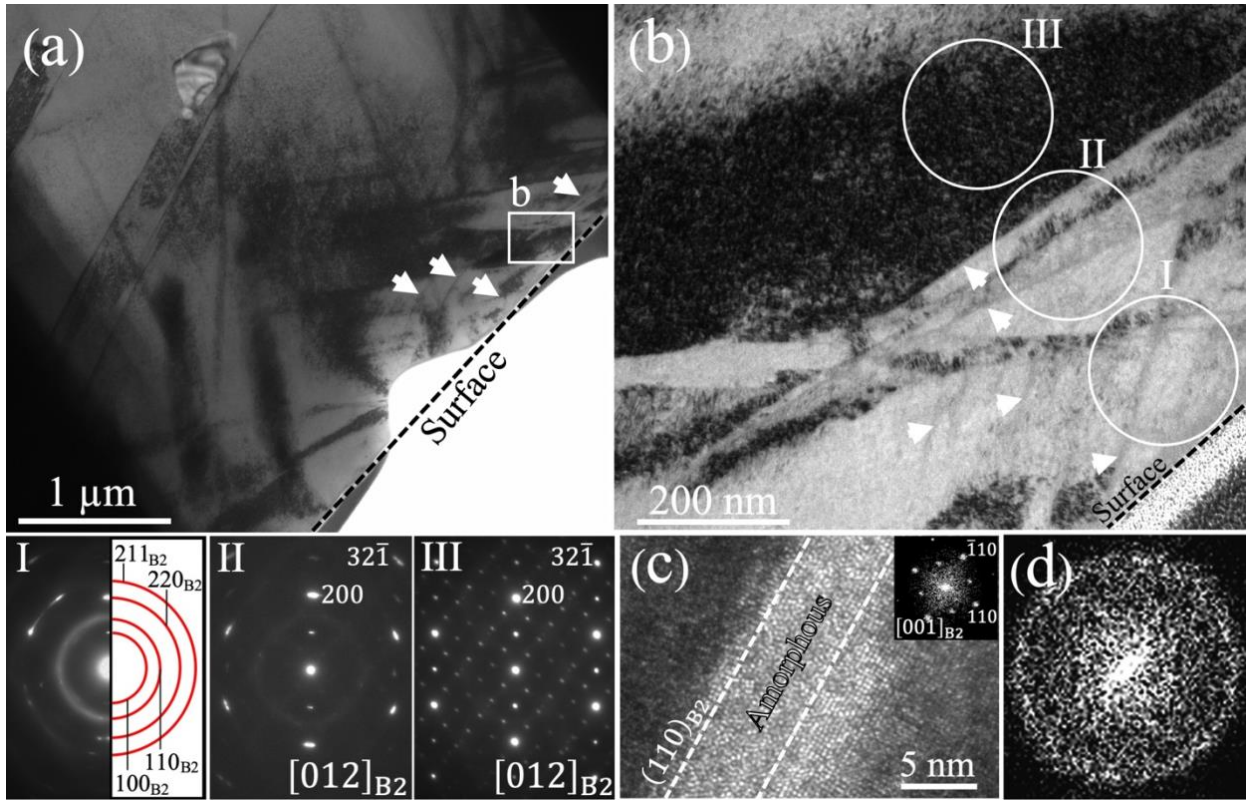

**Fig. 4.** (a) BF-TEM micrograph near the surface of the super-polished $Ni_{55}Ti_{45}$ showing narrow deformation bands (e.g., those marked with white arrows). (b) Magnified BF-TEM micrograph of the region indicated by the white box in (a) showing closely spaced deformation bands (e.g., those marked with white arrows) located close to the surface. (I) SAED pattern taken from an area within 200 nm of the polished surface, showing both strong diffuse rings and distinct B2 spots that correspond to amorphous phase and several B2 nanocrystals being contained within the diffracting volume, respectively, as supported by the schematic overlay of the indexed B2 structure diffraction rings on the right side. (II) SAED pattern taken along the $[012]_{B2}$ zone of a B2 grain within the region 200 – 400 nm below the polished surface, also showing faint diffuse rings from the presence of some amorphous phase and faint $Ni_4Ti_3$ superlattice reflections. (III) SAED pattern taken along the $[012]_{B2}$ zone from a single B2 grain within the region 400 – 600 nm beneath the polished surface, including strong superlattice reflections of $Ni_4Ti_3$ precipitates along $\frac{1}{7}\langle 321 \rangle$. (c) HRTEM micrograph of a deformation band close to surface with interface that lies on $(110)_{B2}$ plane (orientation confirmed by FFT (inset)) and (d) corresponding FFT showing the amorphous structure of the band.

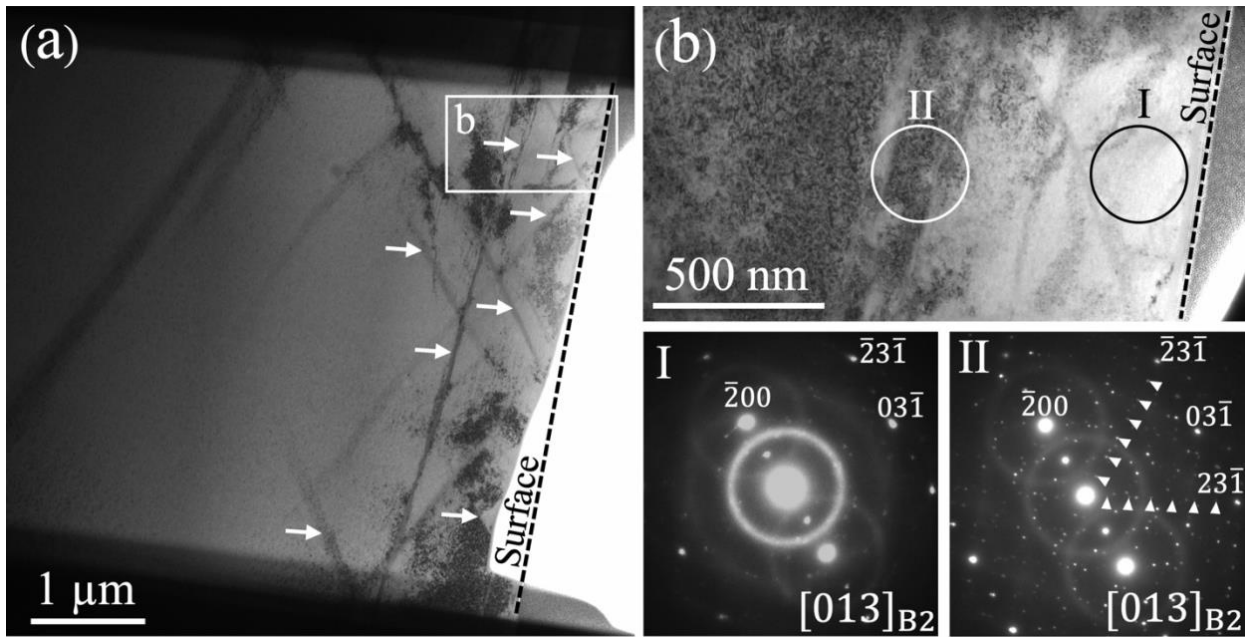

**Fig. 5.** (a) Low magnification BF-TEM of the material beneath the surface of a $Ni_{55}Ti_{45}$ wear track that achieved the runout condition shows deformation bands (e.g., those marked with white arrows). (b) Higher magnification BF-TEM micrograph taken from the region indicated by white box in (a). (I) SAED pattern taken from the material just beneath the wear surface showing clear spot pattern of a single grain along the $[013]_{B2}$ zone, diffuse rings from amorphous bands, and weak superlattice reflections, possibly from nanocrystals and/or $Ni_4Ti_3$ precipitates. (II) SAED pattern taken from a region 800 nm beneath the wear surface shows a B2 grain oriented along the $[013]_{B2}$ zone, and includes intense superlattice reflections from two variants of $Ni_4Ti_3$ precipitates (white arrowheads) and weak diffuse rings (from amorphous bands).

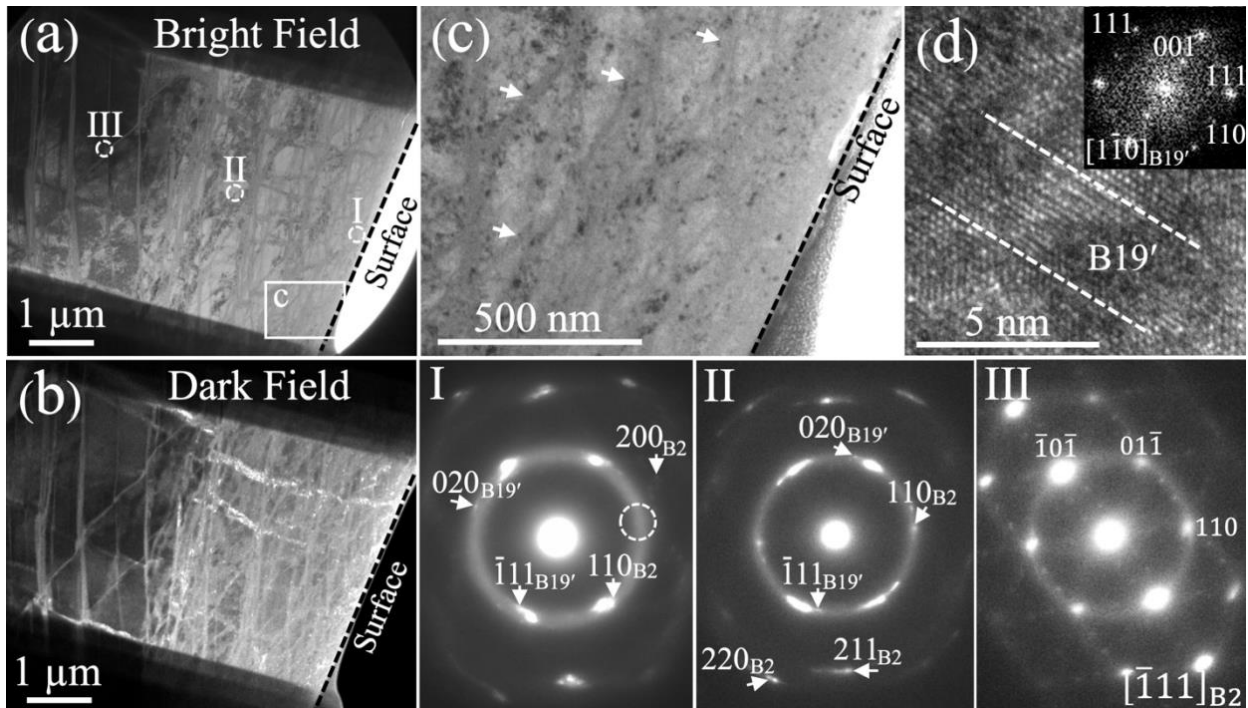

**Fig. 6.** (a) BF-TEM and (b) corresponding DF-TEM micrograph of $Ni_{55}Ti_{45}$ beneath a spalled region of a failed wear track surface. The DF micrograph was imaged using the part of ring pattern marked by the dashed circle on the SAED pattern of (I). (c) Magnified BF-TEM image taken from the region indicated by the white box in (a) contains high density of deformation bands (e.g., those marked with white arrows). (I) SAED pattern taken near the surface (the region within the dashed circle labeled (I) in (a)) contains a diffuse ring and spots that correspond to amorphous bands and B2/B19' nanocrystals, respectively. (II) SAED pattern taken 2.1 – 2.5 µm from surface (the region within the dashed circle labeled (II) in (a)) exhibits a weaker diffuse ring pattern, which suggests less amorphization than at the surface. B2/B19' nanocrystals are also still observed; precipitates were not observed up to 4.5 µm into the sample. (III) SAED pattern taken 4.1 – 4.5 µm from the surface (the region within the dashed circle labeled (III) in (a)) shows a spot pattern of a single B2 grain along the $[\bar{1}11]_{B2}$ zone containing $Ni_4Ti_3$ precipitates, evident from the superlattice reflections. (d) HRTEM micrograph and corresponding FFT pattern (inset) taken close to the surface confirm the B19' structure of some of the nanocrystals.

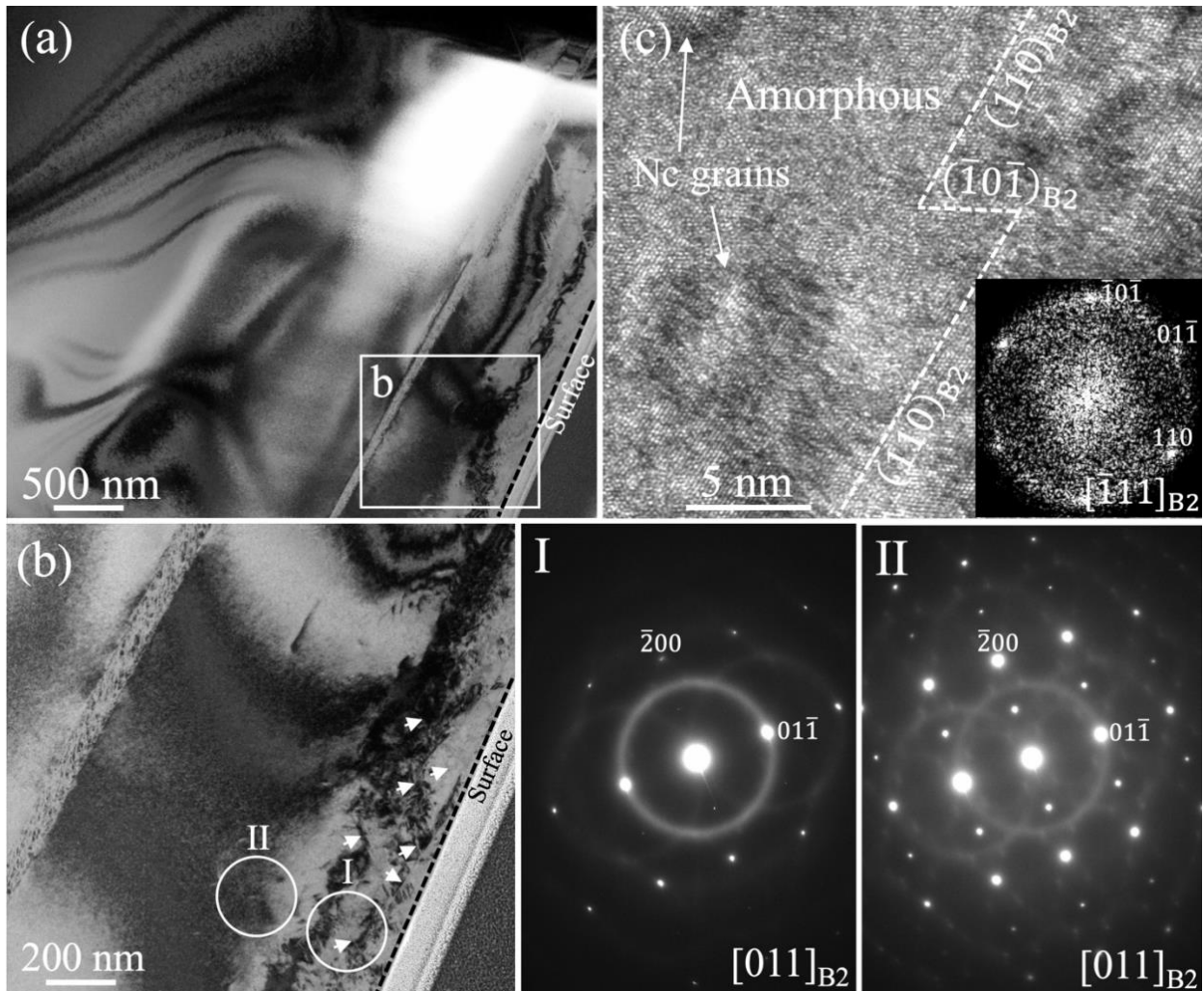

**Fig. 7.** (a) Low and (b) high magnification BF-TEM micrographs of the material beneath the surface of a $Ni_{54}Ti_{45}Hf_1$ super-polished sample show deformation bands (e.g., those marked with white arrows in (b)) close to the surface. (I) SAED pattern taken close to the surface indicated by the white circle in (b), shows spots of a single B2 grain along $[011]_{B2}$ zone, together with a diffuse rings corresponding to amorphous bands. Precipitates were not observed up to 350 nm into the sample. (II) SAED pattern taken 350 – 550 nm from the polished surface (per the ring labeled II in (b)) indicate a B2 grain oriented along the $[011]_{B2}$ zone axis, together with a weak diffuse ring pattern of amorphous bands and $Ni_4Ti_3$ superlattice reflections corresponding to $Ni_4Ti_3$ precipitates. (c) HRTEM micrograph and corresponding FFT (inset) taken close to the surface shows an amorphous deformation band that contains B2 structured nanocrystalline (Nc) grains. The zig-zagged interface of the deformation band follows $(110)_{B2}$ and $(\bar{1}0\bar{1})_{B2}$ planes.

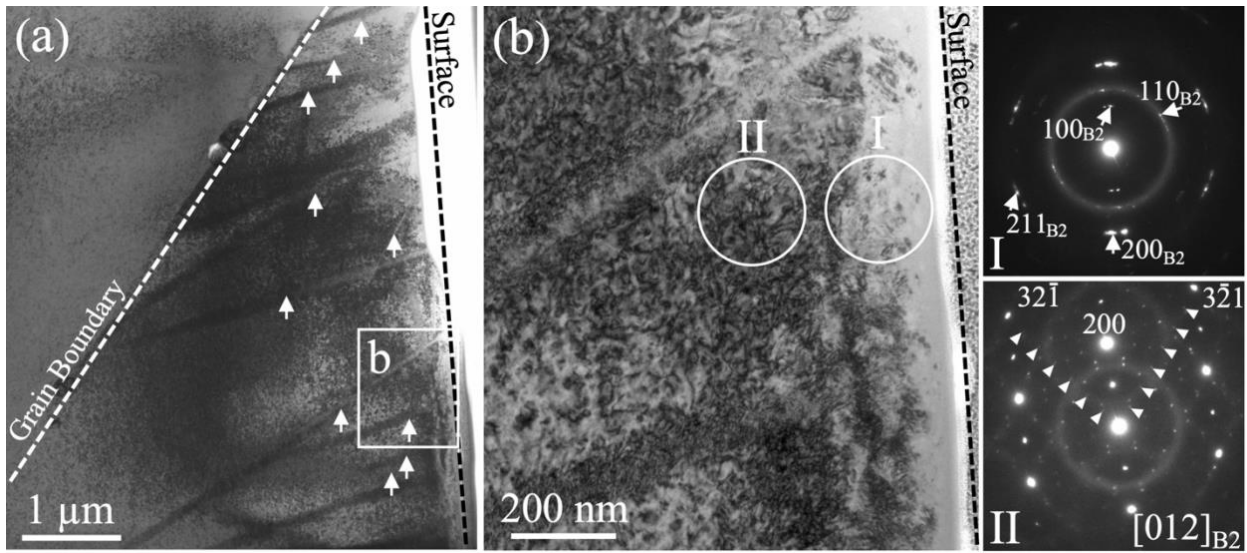

**Fig. 8.** (a) BF-TEM micrograph of the material beneath the surface of a $Ni_{54}Ti_{45}Hf_1$ wear track that achieved the runout condition shows broad deformation bands (e.g., those marked with white arrows) propagating beneath the wear surface. (b) Magnified BF-TEM image indicated by the white box in (a). The material within the circled region marked I consists of a 200 nm amorphous surface layer containing B2 nanocrystals, as confirmed by (I) the corresponding SAED pattern, which exhibits a weak diffuse ring and sharp spots. Precipitates were not observed up to 250 nm into the sample. The material 250 – 450 nm beneath the wear track surface within the circle labeled II in (b), together with (II) the corresponding SAED pattern indicate a single B2 grain oriented along the $[012]_{B2}$ zone axis, containing two variants of $Ni_4Ti_3$ precipitates, as indicated by the superlattice reflections marked with white arrowheads.

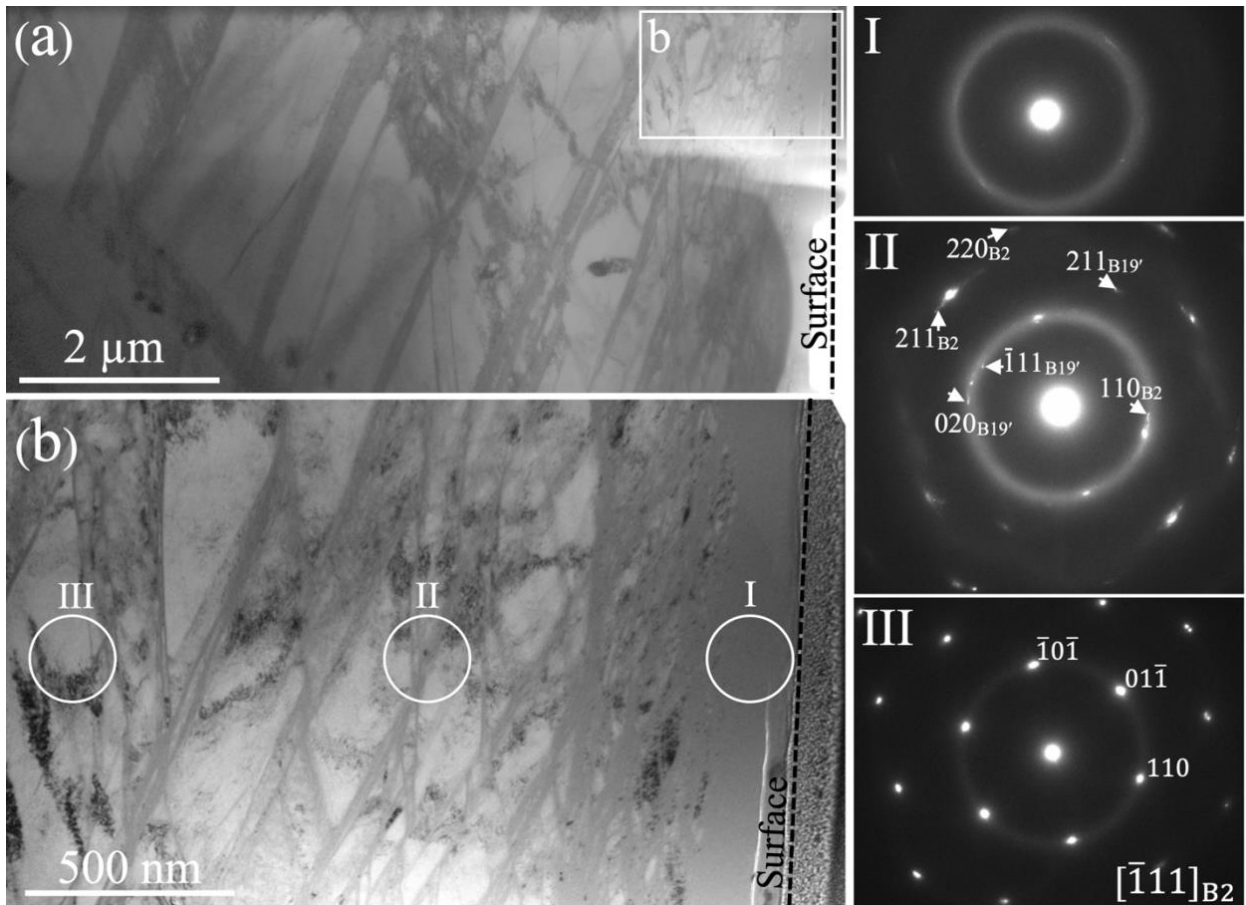

**Fig. 9.** (a) BF-TEM image of the material beneath a spalled region of a failed $Ni_{54}Ti_{45}Hf_1$ spalled wear track shows the prevalence of deformation bands to more than 7 µm depths beneath the sample surface. (b) A higher magnification BF-TEM micrograph from the boxed region in (a), as well as (I-III) SAED patterns taken from the corresponding circled regions in (b). (I) a diffuse ring SAED pattern taken near the surface indicates predominantly amorphous phase. (II) a combined diffuse ring and spot SAED pattern taken 800 nm – 1 µm beneath the surface indicates the presence of both an amorphous phase and B2/B19' nanocrystals. (III) SAED pattern taken 1.6 µm beneath the spalled surface indicates a single B2 grain oriented along the $[\bar{1}11]_{B2}$ zone axis containing dispersed amorphous bands. Evidence for $Ni_4Ti_3$ precipitates was not found anywhere within the FIB sample.

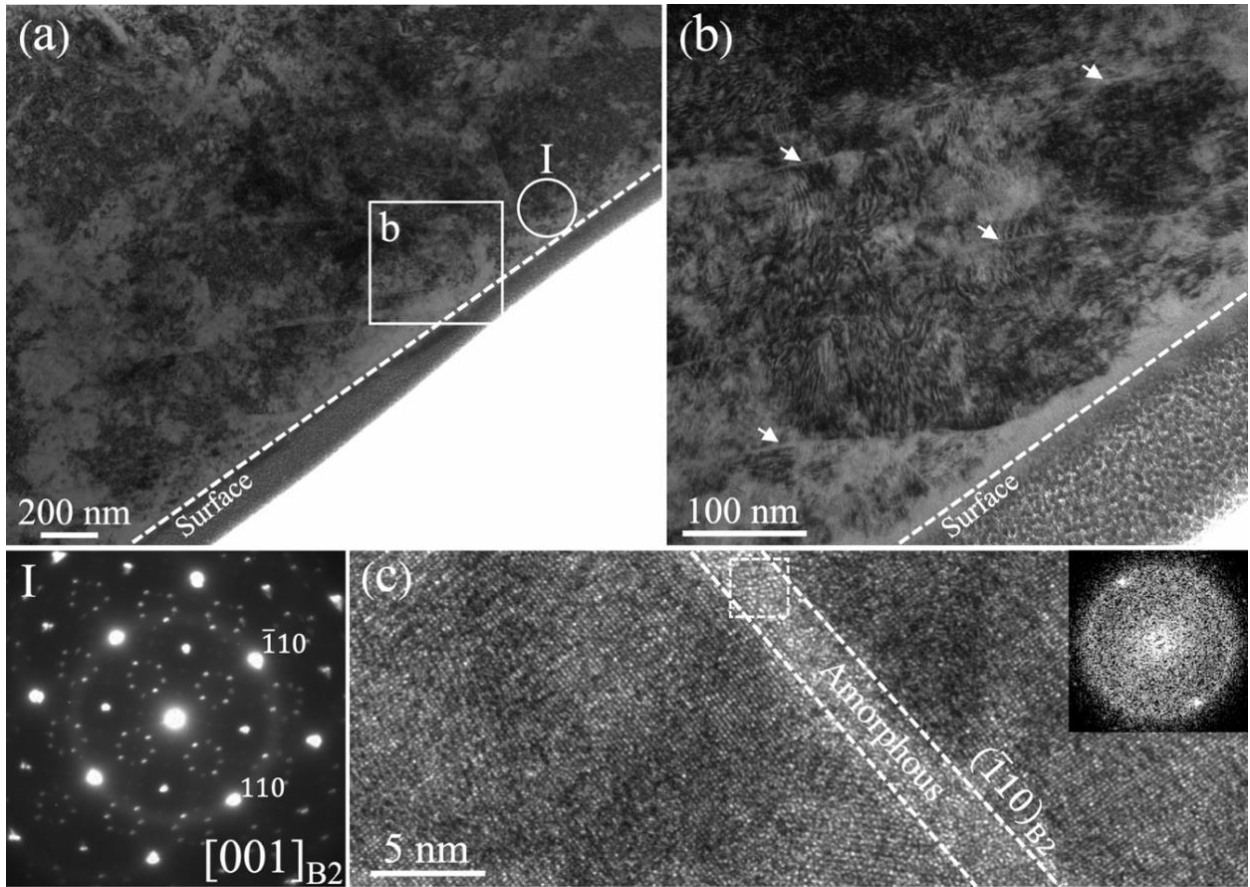

**Fig. 10.** (a) Low and corresponding (b) high magnification BF-TEM micrographs of the material beneath the super-polished (unworn) surface of a $Ni_{56}Ti_{36}Hf_8$ sample. The deformed microstructure contains narrow deformation bands (e.g., those marked with white arrows in (b)). (I) SAED pattern taken near the surface (as circled in (a)) indicates a single B2 grain oriented along the $[001]_{B2}$ zone axis (bright spots), containing amorphous bands (weak diffuse ring pattern), and H-phase/cubic NiTiHf phase precipitates (superlattice reflections). (c) HRTEM image and corresponding FFT (inset) indicates an amorphous deformation band within interfaces along $(\bar{1}10)_{B2}$ planes.

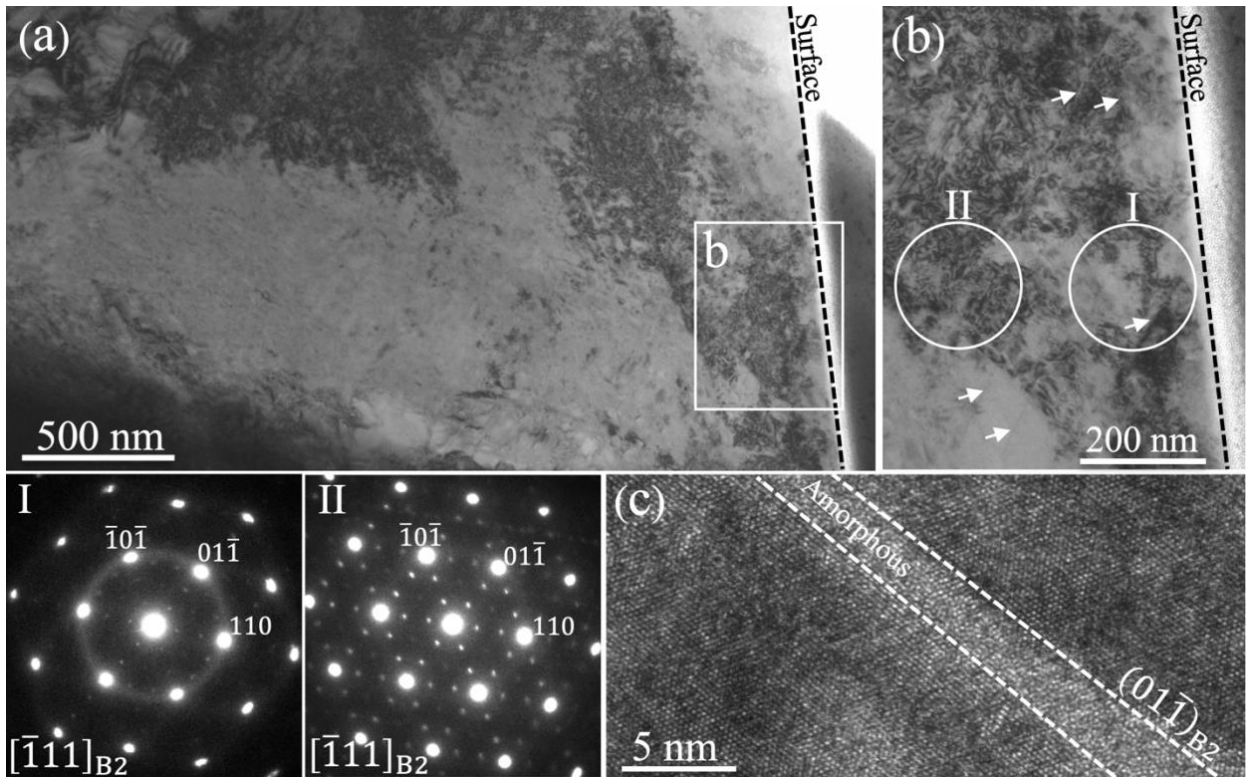

**Fig. 11.** (a) BF-TEM image of the material beneath a runout wear track of a $Ni_{56}Ti_{36}Hf_8$ sample. The corresponding (b) magnified BF-TEM micrograph shows narrow deformation bands (e.g., those marked with white arrows) within the region less than 500 nm from the surface. (I) SAED pattern taken near the surface, as circled in (b), contains an intense spot pattern corresponding to a single grain along $[\bar{1}11]_{B2}$ zone, a diffuse ring originating from thin amorphous deformation bands and weak superlattice reflections from H-phase and cubic NiTiHf precipitates. (II) SAED pattern taken 300 – 500 nm from the surface, as circled in (b), along $[\bar{1}11]_{B2}$ zone of the B2 matrix, contains a very weak diffuse ring from a few amorphous bands and intense superlattice reflections from H-phase and cubic NiTiHf precipitates. (c) HRTEM micrograph taken close to the surface shows a narrow amorphous deformation band with interfaces that follow $(01\bar{1})_{B2}$ planes.

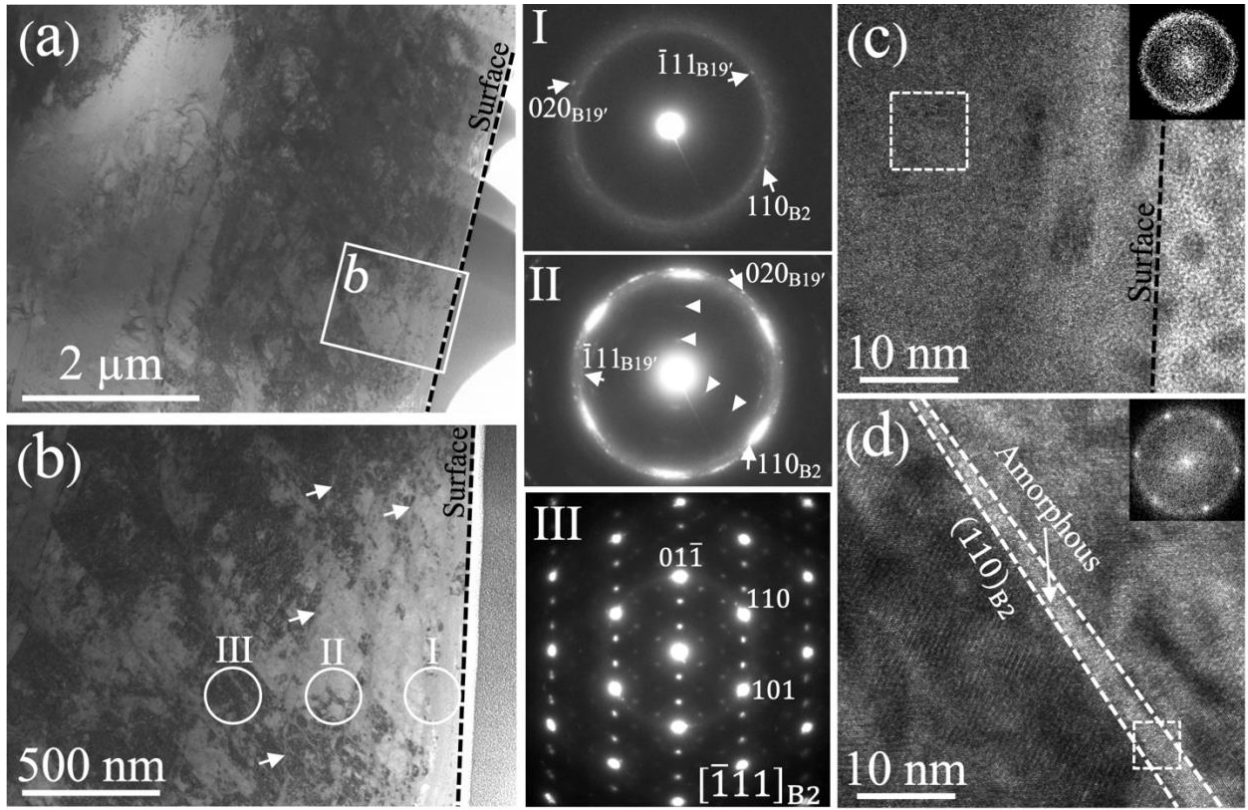

**Fig. 12.** (a) Low and (b) high magnification BF-TEM micrographs of the material beneath a spalled region of a failed wear track of a $Ni_{56}Ti_{36}Hf_8$ sample shows that the heavily damaged region extends 1.5 µm beneath the surface. Narrow deformation bands are marked with white arrows in (b), and regions corresponding to the electron diffraction patterns that are presented are circled. (I) SAED pattern taken near the surface indicates the presence of an amorphous phase (diffuse ring) and both B2 and B19' nanocrystals (spots). (II) SAED pattern taken 200 – 400 nm beneath surface indicates a greater density of B2 and B19' nanocrystalline grains (more intense spots) together with the amorphous phase (diffuse ring pattern). Additional faint reflections such as those indicated by white arrowheads correspond to precipitation. (III) SAED pattern taken 700 – 900 nm from surface indicates a single B2 grain oriented along the $[\bar{1}11]_{B2}$ zone axis, small amounts of amorphous phase (a weak diffuse ring) and H-phase/cubic NiTiHf precipitates (superlattice reflections). This corresponds closely with the typical 3-phase crystalline structure for the alloy in the undeformed super-polished condition (Fig. 10). (c) HRTEM micrograph taken near the surface and corresponding FFT (inset), which confirms a fully amorphous structure very near the wear surface. (d) HRTEM micrograph taken in the primarily crystalline region (III) and corresponding FFT (inset) shows an amorphous deformation band with interfaces that follow $(110)_{B2}$ planes.

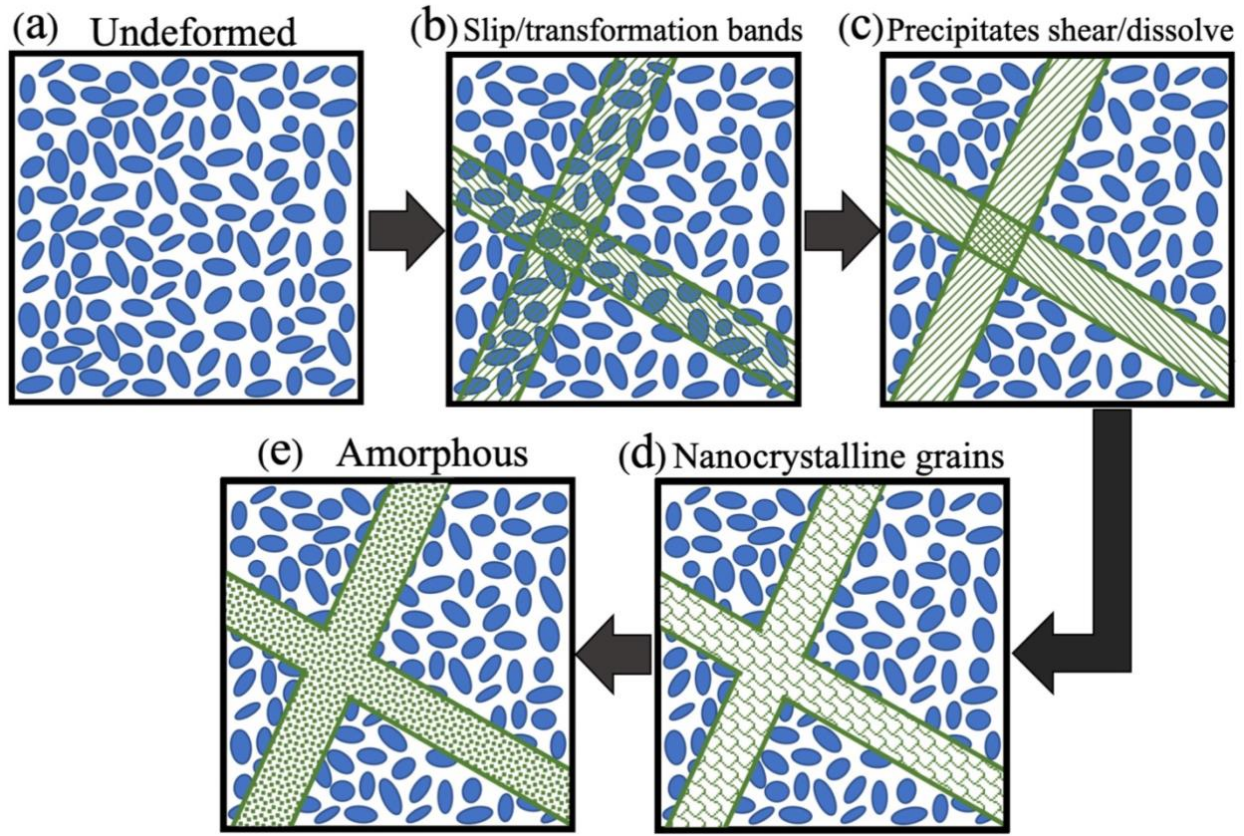

**Fig. 13.** Schematic illustrating the proposed mechanistic sequence that ultimately leads to the formation of amorphous deformation bands within these NiTi-based alloys. (a) Initially (undeformed) the alloys are B2 polycrystals that contain a homogenous distribution of strengthening nanoprecipitates. (b) Initially, localized slip and/or martensitic transformation leads to dislocation accumulation on deformation bands/transformation fronts, then (c) precipitates shear and dissolve within the deformation bands, while (d) dislocation structures rearrange into cells resulting in the formation of nanocrystals, and (e) eventual amorphization. Our results (Figs 5,6,8,9,11,12) indicate that in the most heavily damaged regions, completely amorphized regions/bands are observed, while in less damaged regions, mixtures of these states are observed.